\documentclass[11pt, onecolumn, draftcls]{IEEEtran}

\usepackage{cite}      
\usepackage{graphicx,subcaption} 
\usepackage{amssymb}
\usepackage{verbatim}
\usepackage{amsmath}
\usepackage{url}
\usepackage{amsthm}
\usepackage{enumerate}
\usepackage[pagebackref=true,breaklinks=true,colorlinks,bookmarks=false]{hyperref}
\usepackage{wrapfig}

\usepackage{verbatim}

\makeatletter
\def\th@plain{%
  \thm@notefont{}
  \itshape 
}
\def\th@definition{%
  \thm@notefont{}
  \normalfont 
}
\makeatother

\def\bbtheta{{\mbox{\boldmath $\theta$}}}

\def\bbphi{{\mbox{\boldmath $\phi$}}}

\newif\ifshortref
\shortreffalse
\ifshortref
    \newcommand{\ccite}[1]{\hspace{-0.4em}}
    \newcommand{\arxivref}[1]{$\mbox{#1}$}
\else
    \newcommand{\ccite}{\cite}
    \newcommand{\arxivref}[1]{}    
\fi

\begin{document}
\title{Deploying AI for Signal Processing education: Selected challenges and intriguing opportunities}
\author{Jarvis Haupt, Qin Lu, Yanning Shen, Jia Chen, Yue Dong, Dan McCreary, Mehmet Ak\c{c}akaya and Georgios B. Giannakis\thanks{J. Haupt, M. Ak\c{c}akaya and G. B. Giannakis are with the Dept. of ECE, U. of Minnesota; Q. Lu is with the School of ECE, U. of Georgia; Y. Shen is with Dept. of EECS, UC Irvine; J. Chen and Y. Dong are with the Depts. of ECE and CSE, UC Riverside. The work of Q. Lu is supported by NSF\# 2340049; The work of Y. Shen is supported by NSF \#2207457, 2412484, 2442964 and 2425748; the work of G. B. Giannakis was supported by  NSF grants 2126052, 2220292, 2312547, 2212318, and 2332547; the work of J. Chen was supported by NSF grants 2244480, 2424458, 2431569. Emails: jdhaupt,akcakaya,georgios@umn.edu; qin.lu@uga.edu; yannings@uci.edu; jiac, yue.dong@ucr.edu; dan.mccreary@gmail.com. \arxivref{A version of this paper with a more complete list of references is available on arXiv.}\\
\\
\copyright 2025 IEEE.  Personal use of this material is permitted.  Permission from IEEE must be obtained for all other uses, in any current or future media, including reprinting/republishing this material for advertising or promotional purposes, creating new collective works, for resale or redistribution to servers or lists, or reuse of any copyrighted component of this work in other works.}}

\maketitle

\begin{abstract}
Powerful artificial intelligence (AI) tools that have emerged in recent years—including large language models, automated coding assistants, and advanced image and speech generation technologies—are the result of monumental human achievements. These breakthroughs reflect mastery across multiple technical disciplines and the resolution of significant technological challenges. However, some of the most profound challenges may still lie ahead. These challenges are not purely technical but pertain to the fair and responsible use of AI in ways that genuinely improve the global human condition.
This article explores one promising application aligned with that vision: the use of AI tools to facilitate and enhance education, with a specific focus on signal processing (SP). It presents two interrelated perspectives: identifying and addressing technical limitations, and applying AI tools in practice to improve educational experiences. Primers are provided on several core technical issues that arise when using AI in educational settings, including how to ensure fairness and inclusivity,  handle hallucinated outputs, and achieve efficient use of resources. These and other considerations—such as transparency, explainability, and trustworthiness—are illustrated through the development of an immersive, structured, and reliable “smart textbook.” The article serves as a resource for researchers and educators seeking to advance AI’s role in engineering education.  
\end{abstract}


\IEEEpeerreviewmaketitle

\section{Introduction} 

Academic disciplines grounded in the physical sciences and mathematics tend to be viewed as among the most difficult courses of study for students \cite{novik2022role}, a sentiment to which many educators in these fields can relate.  Characterized by foundational paradigms that are technically rigorous but often abstract, and sometimes lacking concrete physical intuition, STEM-related disciplines can indeed present unique educational challenges.  One potential explanation comes from \emph{cognitive load theory}~\ccite{plass2010cognitive}, which suggests that learning tasks requiring heightened attention tend to demand a higher \emph{cognitive load}, overloading ``short-term'' memory, and ultimately hampering learning effectiveness \cite{de2010cognitive}.  This insight suggests that personalized learning paradigms -- wherein a student can learn at their own pace, potentially aided by an expert ``tutor'' able to answer intermediate questions -- may be among the more appropriate methodologies for STEM education.  Unfortunately, this approach can be largely at odds with more traditional classroom-based lecture-type learning paradigms which must, by design, adhere to a number of more rigid timing and instructor availability constraints.

On the other hand, the rapid pace of technological advancement continues to make available a number of avenues to help address these challenges.  Indeed, the large (and continuously-growing) amount of high-quality digital educational content available online, and the widespread proliferation of computational resources generally, have ushered in entirely new learning pathways with the potential to reduce or eliminate some of the barriers to technical education.  For example, for many challenging technical topics, one can now watch specialized, narrated, guided-tour videos on services such as YouTube.  Highly technical questions can be posed by students and answered by a competent community of like-minded participants on forums such as Stack Exchange or Stack Overflow.  Many topics may even have their own live and active discussion forums on venues like Reddit or Discord.  This is occurring even as resources such as OpenStax serve as repositories for free digital textbooks, and many university educators continue to create and make freely available their own curated educational content.   

Innovative pedagogical strategies such as ``flipped classrooms,'' in which students digest content outside of a classroom setting first, and then participate in an interactive classroom-based problem-solving session afterwards, were perhaps among the earliest forays exploring and evaluating new kinds of technology-fueled learning in the technical disciplines.  To wit, in the past decade numerous studies have established that flipped classrooms can be effective alternatives to the traditional lecture-based learning paradigm~\ccite{akccayir2018flipped}, including in engineering education~\ccite{bishop2013flipped, kerr2015flipped}, and signal processing (SP) in particular \cite{van2013flipping}.  With the recent advances in \emph{artificial intelligence} (AI) tools however, it is becoming increasingly feasible that the next iterations on this general theme are more likely to be revolutionary rather than evolutionary.

\subsection{The promise of artificial intelligence}

Without a doubt, the emergence of AI in the past few years has been one of the most consequential technological advances in generations, disrupting the state-of-the-art on numerous fronts.  In natural language processing, for instance, as more modest rule-based language models used even just five years ago gave way to massively parameterized \emph{attention-based} large language models (LLMs) with millions, billions, or even over a trillion parameters~\ccite{wiki:GPT-4} new waves of automated ``chatbots'' have emerged as powerful multipurpose agents with unprecedented abilities.  These tools are now able to correctly answer a broad range of queries; hold cogent conversations; provide accurate summarizations of articles, web pages, books, and even health records; and even perform on par with human counterparts on standardized evaluations across a broad array of disciplines.  Further, their capabilities continue to advance at an astounding rate, thanks in large part to an active worldwide community spanning scientists, industry, commerce, and hobbyists.

On the audiovisual front, new diffusion-based image and video generation and enhancement tools are now revolutionizing digital content creation. The natural artistic talent that was traditionally a prerequisite for success in many of these disciplines is now being superseded by \emph{prompt engineering} as a neo-artistic means to an aesthetic end.  Indeed, a well-constructed prompt or sequence of prompts now suffices to generate high-resolution images, such as illustrations or graphics for artistic works, or even high-definition videos that have found utility in advertising, entertainment, and even film-making~\ccite{AIFilms}.

The potential of deep learning and AI in general has also been unleashed on the scientific front, to tremendous effect~\ccite{wang2023scientific}. AI can now perform complicated high-level tasks such as proving mathematical theorems~\ccite{yang2024leandojo}\hspace{0.1em}; predicting protein folding structures~\ccite{jumper2021highly}; and automating drug discovery, product development, and clinical trial design~\ccite{mak2024artificial}. The applications are as pervasive and broad as the scientific disciplines themselves, as more and more promising near-term and long-term future applications continue to emerge. For instance, AI is envisioned to be an integral part of the integration, management, and even the SP aspects of \emph{6G} (and beyond) wireless communications systems~\ccite{chowdhury20206g}.   It is clear that AI technologies collectively have the potential to rapidly and simultaneously transform and reshape scientific discovery, the global workforce, and countless other aspects of the human condition.  

In light of this, one can envision a number of exciting potential applications of AI in education and information dissemination. Consider, for example, a trusted resource capable of \emph{automatically} preparing highly-specialized and immersive topic videos, answering highly-specific technical questions across a broad range of topics, engaging in cogent conversations to help dispel uncertainties that a learner may have, or even creating curated (hyper-)personalized text books complete with figures and interactive demonstrations.  This vision may have sounded like science fiction even just a few years ago, but today is entirely plausible given the potential of AI.  

With this as motivation, our overall objective for this article is twofold: first, we identify and discuss several technical challenges in AI that are associated with putting this concept into practice along with the current state-of-the-art on these fronts, and second, we assess the efficacy of current AI tools in creating some of this kind of content in a purely automated and \emph{trustable} way. Our target audience thus includes both researchers who are interested in technical primers on some of the latent challenges present in contemporary AI systems, as well as  practitioners who aim to utilize AI tools to create their own curated and reliable educational content. We provide a brief motivation for each of our aims next, and further elaborate on these aims in the subsequent sections of this article.

\subsection{Outline: Challenges and opportunities}

According to a recent global survey exploring public perception on the trustworthiness of AI, approximately 3 in 5 respondents (61\%) were generally ``wary about trusting AI systems'' \cite{gillespie2023trust}.  While the motivations for the distrust expressed in these survey results could be many, the overarching theme highlights a latent skepticism of the factual reliability of these tools.  This can likely be traced, at least in part and perhaps anecdotally, to the ``black box'' or opaque nature of many of these tools. Typically trained on massive repositories of digital content, and often refined or fine-tuned with user-specific feedback, many commonly-used AI systems offer an obscure glimpse into the semantic meanings of the massive collection of parameters underlying the specific operational mechanisms that turn prompts or queries into useful or actionable outputs. In other words, even though the specific architectures and operational principles inherent to these systems are well-understood, their sheer scale, and the fact that the training processes and the final model parameters that ultimately result from them are comparatively less well-understood, makes it more complicated to derive useful \emph{a posteriori} insights about their operation that can be parlayed into quantifiable metrics of trust. 

However, that is not to say that these shortcomings should merely be tolerated as unavoidable aspects intrinsic to AI.  On the contrary, the \emph{structural} aspects associated with some of these challenges give rise to promising new research directions.  In our exposition here we discuss, in detail, some current advances in dealing with notions of bias, fairness, and privacy; hallucinations; and query efficiency in AI models in Sections~\ref{sec:2}, \ref{sec:3}, and \ref{sec:4}, respectively.  

Trustworthiness issues seem to not bode well for the immediate use of AI as an educational tool, either. However, notwithstanding the aforementioned technical challenges, the use of AI for education in technical disciplines enjoys a unique opportunity that may not be present in other general AI use cases.  Namely, many components of technical education are based fundamentally on foundational (that is, axiomatic or derived) principles which can be proven, verified, or demonstrated. In this way, AI-based educational tools created for and tailored to technical disciplines have, at their disposal, rigorous methods for establishing \emph{fundamental}, or at the very least, \emph{vicarious}, legitimacy and credibility.  This can be achieved, for instance, by appealing to rigorous technical foundations or established analytical results or models, using repeatable experimentation based on sound underlying fundamentals, and endowing themselves with vicarious credibility from reliable referenced sources (or a combination of these). 

While it may be challenging to directly ensure that existing AI models automatically adhere to and utilize these principles, especially on the \emph{consumer} end, and without access to model specifics or the dataset(s) used for training, there are mechanisms by which a user can try to ensure that educational content produced is reputable and reliable.  Among these are explicit generation of \emph{learning graphs} that demonstrate viable learning paths through concepts that build on one another to ultimately achieve certain learning objectives.  Additional methods include automatic generation of bibliographies, perhaps with specific pointers to which parts of the generated content was derived from each source (not unlike how some search engines have begun augmenting their AI-generated search results), restriction to content extracted only from reputable sources (measured, e.g., by their number of citations), and simulation-based modules that \emph{demonstrate} key learning concepts. We highlight several of these in our ``case study'' assessment in Section~\ref{sec:5}. Finally, a few brief concluding remarks are presented in Section~\ref{sec:6}.

\section{Personalized and Fairness-aware AI for Education} \label{sec:2}

Personalized AI in education has gained significant attention as a tool to enhance learning experiences. AI-driven educational technologies leverage machine learning algorithms, data analytics, and adaptive learning strategies to customize instruction to individual student needs. It has been demonstrated that AI-powered personalized learning can markedly improve student engagement and academic performance. AI facilitates adaptive learning environments that adjust content delivery based on students' learning styles, and performance metrics~\ccite{zawacki2019}. Personalized AI tutors can thus provide real-time feedback and customized learning pathways, leading to improved knowledge retention and motivation among learners~\ccite{luckin2018}. Moreover, AI has been shown to enhance inclusivity by catering to diverse student needs, including those with disabilities \cite{holmes2020}. 

In particular, LLMs contribute to personalized education through various applications, including intelligent tutoring systems, automated feedback generation, and AI-driven content creation. Intelligent tutoring systems, such as ChatGPT-based learning assistants, have been shown to deliver personalized explanations and scaffolded problem-solving guidance \cite{koedinger2019}. Recent efforts have also developed AI-driven assessment tools that leverage LLMs to analyze student responses, provide formative feedback, and suggest tailored learning materials \cite{pane2017}. Furthermore, AI-powered analytics are capable of assisting educators in monitoring student progress and adapting instructional strategies \cite{baker2014}.

\subsection{{Fairness-aware AI}}
\looseness=-1
With its advantages granted, implementing AI in education presents challenges. Potential unfairness in AI algorithms arises, as improper training data can result in inequitable learning experiences~\ccite{west2019}. Multiple notions of fairness have been advocated, including the \emph{group fairness} measures of statistical parity,
equal opportunity,  equalized odds,  and the \emph{individual fairness} measures of (un)awareness, and counterfactual fairness~\ccite{kusner2017counterfactual}; see e.g., \cite{dwork2012fairness,hardt2016equality} for definitions of statistical parity and equal opportunity. Formally speaking, given a binary sensitive attribute $s$ and two sensitive subgroups $\mathcal{S}_0, \mathcal{S}_1$, the measures of statistical parity \cite{dwork2012fairness} and equal opportunity \cite{hardt2016equality} are given, respectively, by
\begin{align}
\Delta_{SP}:=&|P(\hat{y}=1 \mid s=0)-P(\hat{y}=1 \mid s=1)| \label{criterion:sp}  \\ 
\Delta_{EO}:=&|P(\hat{y}=1 \mid y=1, s=0)- P(\hat{y}=1 \mid y=1, s=1)| \label{criterion:eo}
\end{align}
where $y$ denotes the ground-truth label and $\hat{y}$ denotes the predicted label. In particular, $\Delta_{SP}$   equalizes the positive rate in two sensitive groups, and thus achieves an equal distribution in the long run, while 
$\Delta_{EO}$ considers the ground-truth label, and aims to achieve similar accuracy across different sensitive groups. 

Fairness and bias can lead to serious consequences when adopting AI to education systems. For example, suppose an AI-powered academic advising tool is used to recommend engineering courses and career paths to students based on their academic performance, interests, and historical data. However, the AI tool was trained on historical data, where female and minority students were underrepresented in certain engineering fields, including electrical or mechanical engineering. As a result, AI may rarely recommend those fields to women, instead steering them toward disciplines such as environmental or biomedical engineering, thus reinforcing existing gender and diversity gaps in the profession.

A potential remedy for this issue is to introduce fairness-aware techniques -- AI tools should be explicitly designed to recognize and correct for systemic underrepresentation. For example, if the system sees that a female student has strong math and design skills, it can recommend mechanical or electrical engineering, even if past data would typically suggest otherwise. Furthermore, interest-driven matching could also facilitate fairness in decision-making of AI-guided education systems. Specifically, rather than relying heavily on demographic or historical data, the AI tool can focus on individual student preferences, goals, and learning styles to suggest personalized pathways. In addition,  bias correction techniques could be introduced, including i) preprocessing, where the training data is carefully selected to avoid bias due to historical skewness in training data distributions; ii) in-processing, by introducing fairness-aware objective functions or neural network architectures to incorporate fairness during model training; and iii) post-processing, by monitoring the model output and adjusting for skewed patterns to align with individual preference and need. 

It is also worth noting that reliance on AI may lead to reduced human interaction, which is essential for social and emotional learning~\ccite{selwyn2020}. Addressing these challenges requires transparent AI governance and continuous monitoring of AI-driven educational systems. Hence, future LLMs in education can focus on advances in AI personalization, explainability, and human-AI collaboration. One potentially viable approach advocates \emph{hybrid} educational models, where LLMs support rather than replace educators~\ccite{luckin2018}. 

\subsection{Graph-based interpretable AI for education}

AI tools for education will aim to enhance model interpretability, but also mitigate biases, and ensure fairness and inclusivity. Graph-based models present a promising venue for the advancement of personalized AI in education; see, e.g., \cite{fan2024graph}~\ccite{li2024graph}. These models leverage the interconnected nature of learning concepts by representing knowledge as nodes and their relationships as edges, allowing for more nuanced and context-aware recommendations. Using graphs to map student learning pathways, AI systems can better understand individual progress, identify conceptual gaps, and suggest personalized learning trajectories. Additionally, graph neural networks (GNNs) can enhance adaptive learning by dynamically adjusting content based on a student’s evolving knowledge graph, leading to a more efficient and targeted educational experience~\ccite{Wu2020} \cite{Kipf2017}. Future research should focus on optimizing graph-based models for interpretability, fairness, and integration with existing AI-driven educational platforms~\ccite{zhou2020}. This will enable incorporation of various fairness-aware and explainability techniques for learning over graphs, including but not limited to graph data augmentation, fair graph attention, fairness-aware graph SP, and graph generative models, see e.g.,\cite{ying2019gnnexplainer}~\ccite{kose2023demystifying,kose2021fair,kose2022fairnorm,kose2022fast,kose2023fairgat,kose2023fairnessfilt,kose2024fairwire}. Employing (graph) generative AI to train models for education is also a fruitful direction because graphs provide interpretability in encoding causal relationships, but they can also capture prior knowledge. Notwithstanging, graph generative models can generate synthetic graph data mimicking those drawn from the distribution of real-world graphs that can be used for data-hungry LLM training. Graphs can also alleviate privacy concerns associated with educational data sharing~\ccite{zhu2019dp,buehler2023generative,ying2009graph,zhao2023graphgpt}.

Graph-based causal inference has the potential to enhance curriculum adaptability, optimize assessment strategies, and refine LLM-generated feedback. This is a promising area, where SP education can significantly benefit from AI; see e.g., \cite{gui2023challenge}~\ccite{ma2024causal,liu2024large}. Causal inference allows educators and AI tools to go beyond second-order correlations, and further capture the generally nonlinear underlying cause-effect relationships between various educational interventions and student outcomes. By leveraging methods such as propensity score matching, instrumental variables, and causal graphs, AI-driven educational platforms can more accurately determine the impact of various teaching strategies and adapt learning materials accordingly; see e.g., \cite{pearl2009}~\ccite{imbens2015}. For instance, it can help identify whether changes in instructional design (such as incorporating more hands-on labs or simulations) directly improve student performance in SP courses, or whether a student's engagement with certain types of feedback leads to improvement in conceptual understanding. Ultimately, introducing graph and causal reasoning to LLMs has the potential to revolutionize adaptive learning~\ccite{giannakis2018topology,shen2017kernel}, making AI-driven education more transparent (by showing why certain feedback or content is provided), and interpretable (by making the logic behind personalized recommendations understandable to both students and educators).

\section{Tolerating Hallucination in Large Language Models}
\label{sec:3}

Using generative models, particularly LLMs, can markedly advance the impact of AI to  education~\ccite{zhang2024simulating,xu2024large}.  However, tayloring LLMs for education also presents substantial risks due to their tendency to produce hallucinations, or text that appears credible but is factually incorrect or entirely fabricated~\ccite{maynez-etal-2020-faithfulness}. This issue is particularly concerning in educational contexts, as misleading information can appear authoritative, potentially confusing students and reinforcing misconceptions~\ccite{perkovic2024hallucinations}. Such inaccuracies are especially harmful in fields that depend on precise and verifiable knowledge, such as SP, as well as other disciplines in science, engineering, medicine, and law~\ccite{tang2024prioritizing}. Furthermore, social sciences, where interpretation, critical analysis, and argumentation are fundamental, are also susceptible to misinformation generated by LLMs~\ccite{rossi2024problems}. For example, an early version of GPT-4, when prompted to provide 10 research papers on hallucination reduction, produced a list in which nine out of ten references were non-existent \cite{dong2023future}. This demonstrates the potential for LLMs to fabricate academic sources, misleading students and researchers who may not verify citations carefully.

While limited research efforts have explored LLMs' educational use~\ccite{su2023unlocking}, the impact of hallucinations in learning settings is underexplored~\ccite{espark_ai_education}. In this section, we will survey the literature on hallucination detection and reduction, focusing on methods that enhance the factual accuracy of model-generated content. We will further outline perspectives on adapting these methods to heighten their suitability for SP education. We categorize existing approaches into three paradigms: fine-tuning, retrieval-augmented generation, and decoding-based approaches.

To strengthen the educational impact of this survey, we emphasize how different hallucination types and mitigation strategies affect student learning specifically in SP. We compare these methods in terms of their ability to handle equations, definitions, algorithmic consistency, and logical reasoning, all of which are essential for SP pedagogy. The aim is to inform both developers of LLM-based tools and educators who seek to responsibly integrate these models into technical instruction.

In natural language processing, hallucination refers to having generated text that is either nonsensical or unfaithful to the given source content. In traditional natural language generation, hallucinations are categorized into intrinsic and extrinsic types~\ccite{filippova-2020-controlled, maynez-etal-2020-faithfulness}. This definition is widely used in tasks such as text summarization~\ccite{maynez-etal-2020-faithfulness,cao2021hallucinated} and closed-domain question answering~\ccite{longpre2021entity}, where models are expected to generate outputs based on a given input, e.g., summarizing a section from a SP textbook. If a model produces content that contradicts or deviates from the input, it is classified as an \textit{intrinsic hallucination}. Conversely, if the generated content contradicts established world knowledge, it is considered an \textit{extrinsic hallucination}. In both cases, the content cannot be verified against either the provided input (intrinsic) or external factual sources (extrinsic)~\ccite{2023halluciation}. A recent framework distinguishes between \textit{factuality hallucinations} and \textit{faithfulness hallucinations} \cite{10.1145/3703155}. A factuality hallucination emerges when the generated content is inconsistent with verifiable real-world facts, regardless of whether it aligns with the provided input. This type of hallucination is particularly problematic in domains requiring high factual accuracy. For instance, if an LLM generates an explanation of a SP algorithm but includes an incorrect equation or misrepresents a fundamental principle, it constitutes a factuality hallucination. Because SP relies on precise, mathematically grounded, and verifiable information, even minor inaccuracies can introduce critical misunderstandings that hinder both theoretical learning and practical application.

On the other hand, a faithfulness hallucination refers to cases where the generated content lacks self-consistency or coherence \cite{10.1145/3703155}. This can occur when an LLM produces logically inconsistent statements, fails to maintain coherence across multiple generated responses, or presents conflicting claims within the same output. Faithfulness hallucinations are particularly concerning in long-form text generation and reasoning tasks, where maintaining internal logical consistency is critical. In SP education, this issue can manifest when a model provides incoherent explanations of algorithms, incorrectly links concepts across different problems, or generates inconsistent interpretations of signal transformations, leading to confusion rather than clarity.

Given the major role hallucinations can play in SP education, their reduction is of paramount importance, and can be organized in three forms: (i) fine-tuning, which directly modifies model parameters; (ii) retrieval-augmented generation (RAG), which grounds LLM outputs in external knowledge sources (e.g., textbooks); and (iii) prompt optimization, which mitigates inference-time without altering model parameters.

\subsection{Fine-tuning LLMs}
A core research effort in mitigating hallucination centers on the limitations inherent to the model architectures, novel model designs, training objectives, and data specifically tailored to address these flaws. Fine-tuning strategies address limitations in LLM architectures and training objectives that contribute to hallucinations. To overcome issues with unidirectional representation, a bidirectional autoregressive model, namely BATGPT, has been developed to enhance contextual understanding \cite{li2023batgptbidirectionalautoregessivetalker}. Likewise, encoder-decoder models have merits in optimizing context usage~\cite{liu-etal-2024-lost}. 

Pretraining objective refinements also help to mitigate hallucinations; cf.  \emph{In-Context Pretraining}, where related documents are concatenated in training to promote logical coherence across document boundaries~\cite{shi2024incontext}. Fine-tuning has also been advocated to address hallucinations arising from belief misalignment, where models prioritize user approval (sycophancy) over factual accuracy. Synthetic-data intervention was introduced for fine-tuning LLMs on data where claims are independent of user opinions, thus reducing sycophantic tendencies \cite{wei2024simplesyntheticdatareduces}. 

\subsection{Retrieval-augmented generation LLMs}
Unlike parameter updates that align a model's internal beliefs with factuality, retrieval-augmented generation (RAG) mitigates hallucinations by grounding language model outputs to external knowledge sources, such as textbooks or research papers. This reduces reliance on internal parametric knowledge and enhances factual accuracy and consistency—particularly valuable in SP education, where theoretical and mathematical correctness is crucial.

The simplest form of RAG with LLMs follows a retrieve-and-append approach. By appending retrieved content as input context, this approach ensures responses are grounded by authoritative sources. In-Context retrieval augmented language models (RALM) were introduced in \cite{ram-etal-2023-context}, which directly prepends the retrieved documents to the LLM’s input, and thus improves factual consistency. Beyond text-based retrieval, structured knowledge graphs \cite{baek-etal-2023-knowledge}~\ccite{wen-etal-2024-mindmap} have been leveraged to enhance reasoning and factual accuracy.

In general, RAG can effectively reduce errors due to factual hallucinations by ensuring LLMs reference accurate formulas and theoretical explanations. But to mitigate faithfulness in hallucinations requires maintaining a consistent reasoning chain. In order to strengthen logical consistency, chain-of-thought guided retrieval (e.g.,  \cite{he2022rethinkingretrievalfaithfullarge}~\ccite{trivedi-etal-2023-interleaving,yao2022react}) integrates external knowledge per reasoning step. Moreover, iterative frameworks (e.g., \cite{shao-etal-2023-enhancing}~\ccite{wei2024simplesyntheticdatareduces}) dynamically refine retrieval, while active RAG re-queries low-confidence outputs \cite{jiang-etal-2023-active}. 

\subsection{Decoding-based approaches}

Beyond the model architecture and the knowledge it utilizes -- whether parametric internal knowledge or an external knowledge base, a third critical factor contributing to hallucinations is the decoding algorithm, which determines how tokens are selected during generation. For instance, temperature-based sampling in natural language processing introduces randomness by adjusting sampling diversity: a higher temperature increases variability but also raises the likelihood of hallucinations~\ccite{banerjee2024llmshallucinateneedlive}. Consequently, inference-time strategies for hallucination reduction aim to refine decoding processes to enhance factual accuracy without requiring model retraining or external contexts.

\begin{wrapfigure}{r}{0.7\textwidth}
    \centering
    \vspace{-2em}
    \includegraphics[width=0.99\linewidth]{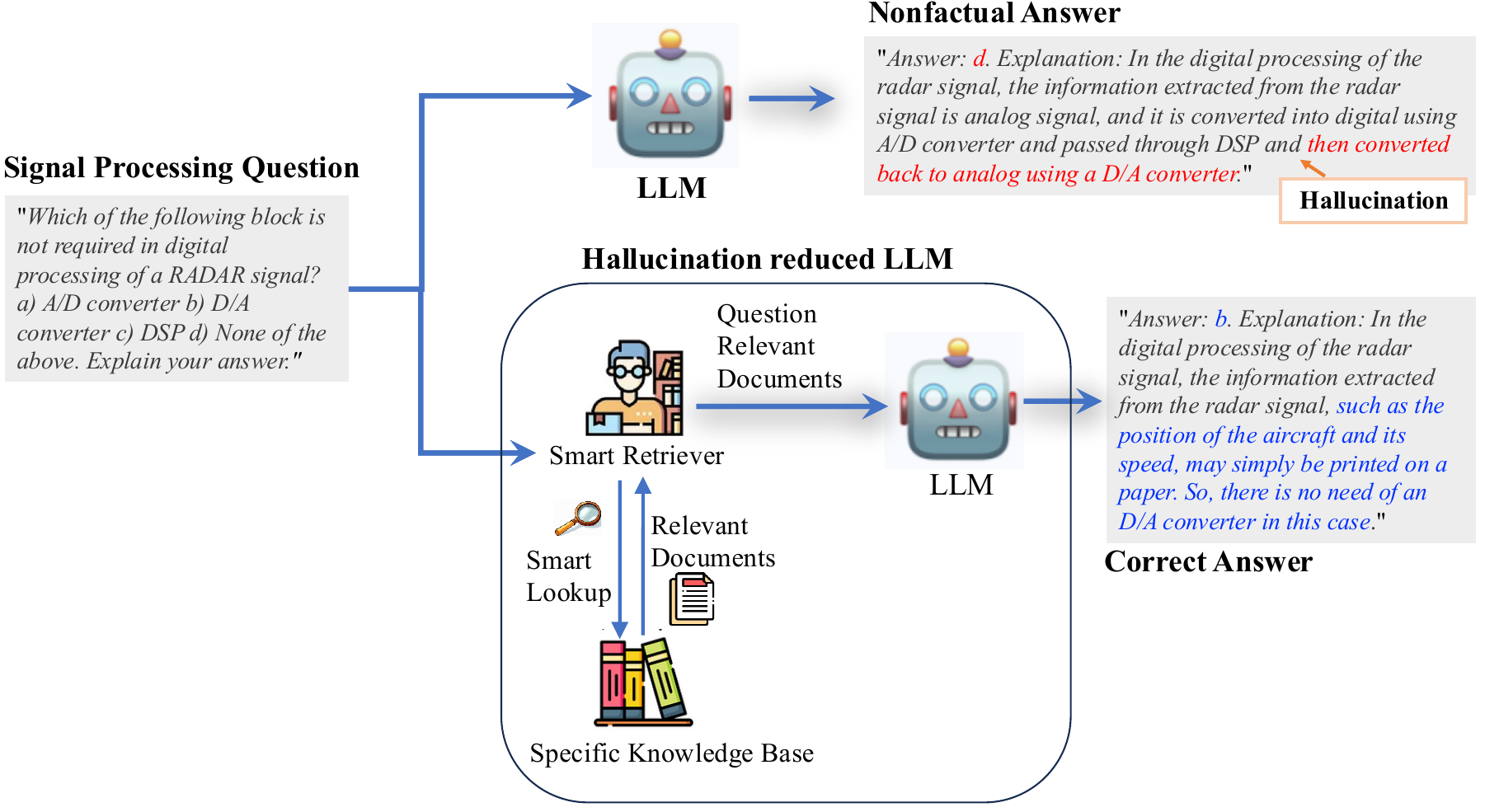}
    \caption{\small Reducing hallucinations of LLMs for SP education.}
    \vspace{-2em}
    \label{fig:LLMHall}
\end{wrapfigure}

Faithfulness-enhanced decoding reduces hallucinations by ensuring that the generated content remains consistent with the input context, and follows a logically coherent reasoning process. These methods are particularly valuable in domains requiring precise alignment with provided information, such as SP, where technical explanations must be both accurate and logically sound. 

A key challenge in hallucination reduction is context misalignment, where models fail to attribute information correctly. Context-aware decoding addresses this by modifying the output distribution to reinforce contextual grounding. This is done by contrastively adjusting token probabilities to prioritize a given context while minimizing reliance on prior knowledge \cite{shi-etal-2024-trusting}. 

Beyond in-place adjustments, post-editing strategies can further enhance context consistency. In a research-then-revise framework,  the model generates initial responses, retrieves supporting evidence, and refines outputs to resolve inconsistencies \cite{gao-etal-2023-rarr}. An entity-level hallucination detection system that identifies hallucinations at both the sentence and entity levels before applying corrections, was developed in \cite{lei2023chainnaturallanguageinference}. To address softmax limitations in maintaining both diversity and faithfulness, a mixture-of-softmax uses multiple hidden states to refine probability distributions \cite{10.5555/3454287.3454806}.

Logical inconsistencies present yet another major source of hallucinations, particularly in multi-step reasoning tasks. In general, using traditional chain-of-thought (CoT) reasoning enhances intermediate reasoning coherence,  but may still produce misleading rationales. This was addressed  with self-consistent CoT, which uses contrastive decoding and a counterfactual reasoning objective \cite{branco-etal-2021-shortcutted}~\ccite{wang-etal-2023-scott} to eliminate reasoning shortcuts. Counterfactual and causal preference optimization aligns models toward valid reasoning chains while avoiding misleading counterfactuals \cite{paul-etal-2024-making}.  A structured approach with symbolic chain-of-thought (SymbCoT) was developed in \cite{xu-etal-2024-faithful}, which translates natural language reasoning into symbolic representations, and thus enforces step-by-step logical consistency.

\subsection{Summary}
In this section, we have introduced the notion of LLM hallucinations, and provided a brief review of state-of-the-art methods to detect and reduce hallucination in three categories: fine-tuning, retrieval-augmented generation, and encoding-based approach. These methods can be adapted to improve their suitability for SP education; see e.g., Fig. \ref{fig:LLMHall}, where a student in a SP class can get the correct answer of their question through hallucination-reduced LLMs. 

\section{Query-efficient design of AI for SP education} \label{sec:4}

While AI tools offer great potential to revolutionize SP education, judicious design is needed to deliver the desired goals. For example,  AI-enabled prediction of students' performance~\ccite{albahli2024efficient}, design of intelligent tutoring system, and detection of plagiarism in tests all hinge on the hyperparameters of the methods employed.  In education content delivery, an instructor seeks to determine the optimal length for video lessons so as to maximize student engagement measured by watch time and quiz scores. As another example, to design SP course materials using LLMs, judicious selection of the prompt is needed to craft instructions for LLMs to follow~\ccite{liu2023pre}. All these applications can be abstracted as the optimization problem $\bbtheta_* ={\arg \max}_{\bbtheta \in \Theta} \ r(\bbtheta)$, where $\Theta$ is the feasible set for $d$-dimensional optimization variables $\bbtheta$, and $r$ is the objective function that assesses the evaluation results of a specific education strategy. Unlike other engineering tasks, $r$ here is a ``black-box” without analytic expression, which prevents one to adopt the conventional gradient-based solvers. Further, each evaluation of the design can be extremely time- and resource-costly. In the prompt optimization task, the process of learning a good prompt requires interactions with the LLM and evaluating its responses, which incurs high cost, thus rendering naive grid search infeasible. To address this bandit optimization task in a sample-efficient manner, we will advocate SP education focused on the Bayesian optimization (BO) framework~\cite{garnett2023bayesian} that actively selects the strategy to find the optimal design with as few queries as possible. To this end, we will first outline the fundamentals of BO, following which we will delineate the prompt optimization task as a case study of the BO-based design approach. Lastly, we will outline emerging topics to enhance the potential of BO of various design tasks for SP education.

\subsection{BO basics}
\begin{wrapfigure}{r}{0.5\textwidth}
    \centering
    \vspace{-1.5em}
    \includegraphics[width=0.85\linewidth]{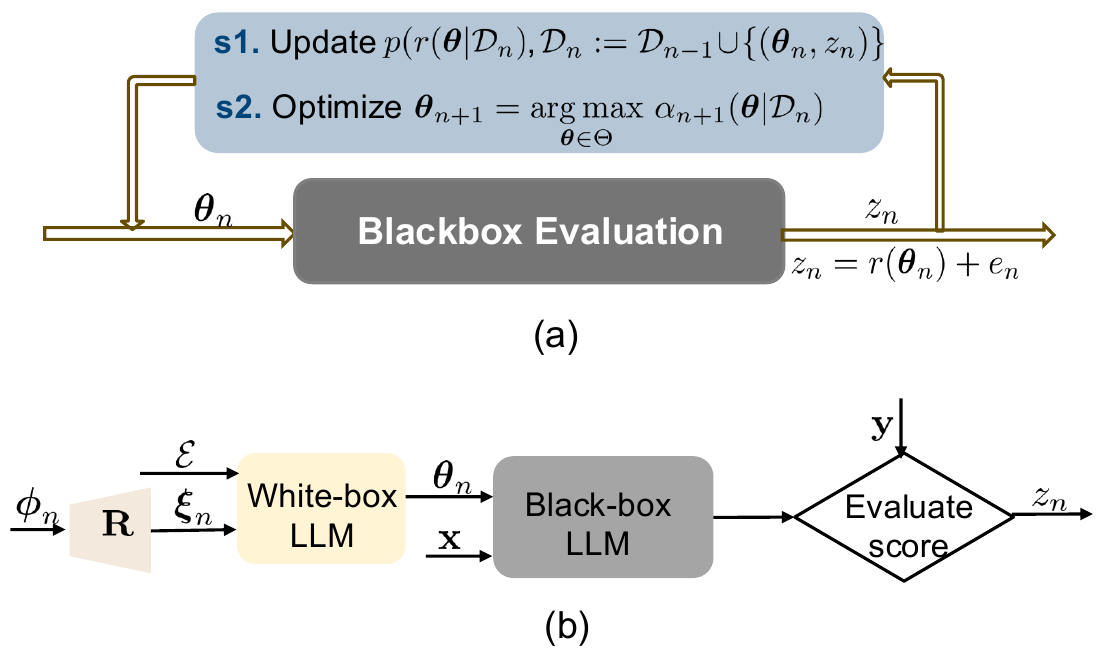}
    \caption{\small BO for design of AI-based SP education with (a) Schematic diagram of BO, and (b) objective function evaluation in prompt optimization, which can be used to design intelligent SP textbooks using LLMs (cf. Sec.~\ref{sec:5}). }
    \vspace{-2em}
    \label{fig:BO}
\end{wrapfigure}
In a nutshell, BO relies on a Bayesian surrogate model for the sought black-box objective $r(\bbtheta)$ to {\it actively} select the query points sequentially~\cite{garnett2023bayesian}. Let $\mathcal{D}_n:=\{ \bbtheta_{n'}, z_{n'}\}_{n'=1}^n$ be the set of  evaluated input-output pairs up to slot $n$, where $z_{n'}$ is the noisy version of $r(\bbtheta_{n'})$. The selection of $\bbtheta_{n+1}$ is implemented iteratively in two steps: 
(i) Find  $p(r(\bbtheta)|\mathcal{D}_n)$ using a chosen surrogate model; and,  
(ii) select $\bbtheta_{n+1}\! =\! {\arg\max}_{\bbtheta\in\Theta}  \ \alpha (\bbtheta|\mathcal{D}_n)$,  where the acquisition function (AF) $\alpha$, usually chosen to have closed form, is designed based on $p(\! r(\bbtheta)|\mathcal{D}_n)$ to balance {\it exploration} with {\it exploitation}; see also Fig.~\ref{fig:BO} (a) for an overview of BO iterations.

\subsubsection{Surrogate models}
The surrogate model plays a performance-critical role in BO. A `good' surrogate model should be able to reason the uncertainty of the learning objective using a limited number of labeled samples. The Gaussian process (GP) is the most widely used surrogate model in BO, because of its uncertainty quantification and sample efficiency. The GP model further yields closed-form expression for the posterior probability density function (pdf). In this context, the unknown learning function is postulated with a GP prior as $r \sim\mathcal{GP}(0,\kappa(\bbtheta,\bbtheta'))$, where $\kappa(\cdot,\cdot)$ is a kernel (covariance) function measuring pairwise similarity of  any two inputs $\bbtheta$ and $\bbtheta'$. This GP prior induces a joint Gaussian pdf for any $n$ function evaluations ${\bf r}_n := [r(\bbtheta_1),\ldots, r(\bbtheta_n)]^\top$  $p({\bf r}_n| \Theta_n) = \mathcal{N} ({\bf r}_n ; {\bf 0}_n, {\bf K}_n)$, where  $[{\bf K}_n]_{i,j} = {\rm cov} (r(\bbtheta_i), r(\bbtheta_j)):=\kappa(\bbtheta_i, \bbtheta_j)$. Value $r(\bbtheta_{n'})$ is linked with the noisy $z_{n'}$ via the Gaussian per-datum {\it conditional} likelihood $p(z_{n'}|r(\bbtheta_{n'})) = {\cal N}(z_{n'};r(\bbtheta_{n'}), \sigma_\epsilon^2)$ ($\sigma_\epsilon^2$ is the noise variance), which is conditionally independent across samples. Then Bayes' rule yields the function posterior pdf as $p(r(\bbtheta)|\mathcal{D}_n) = \mathcal{N}(r(\bbtheta); \mu_{n}(\bbtheta), \sigma_{n}^2(\bbtheta))$ where the mean $\mu_{n}(\bbtheta)$ and variance $\sigma_{n}^2(\bbtheta)$, depending on the kernel function and ${\cal D}_n$, have closed-form expressions~\ccite{Rasmussen2006gaussian}. In addition to GPs, the surrogate model can alternatively be represented by Bayesian neural networks (NNs), random forests, as well as tree-structured Parzen estimators~\cite{garnett2023bayesian}.

\subsubsection{AF design and optimization}
Having available the function posterior pdf that offers the uncertainty values, the next query point $\bbtheta_{t+1}$ can be readily selected using off-the-shelf AFs that strike a balance between exploration and exploitation. The workhorse AF for BO is based on the so-termed expected improvement (EI), that chooses the next query point by maximizing the EI over the current best objective value $r_t^*$; that is, $\bbtheta_{t+1} = \arg\max_{\bbtheta\in \Theta}\ \mathbb{E}[(r(\bbtheta)-r_t^*)^{+}|{\cal D}_t]$. Besides EI, other widely used AFs include the upper confidence bound, Thompson sampling, and entropy search~\cite{garnett2023bayesian}. Given its explicit expression, the AF can be optimized via the available gradient-based or evolutionary search approaches. 

\subsection{Case study on prompt optimization}
Here, we will outline the application of BO to the prompt optimization task, which has gained popularity thanks to its ability to improve the performance of LLMs without parameter fine-tuning~\ccite{liu2023pre}, as discussed in Sect.~\ref{sec:5}. For example, to design an intelligent SP textbook using LLMs as elaborated in the next section, proper prompts have to be designed to deliver the desired performance. While a few early attempts have been made towards prompt engineering in white-box/open-source LLMs, the focus here is on the more powerful black-box LLMs (e.g., ChatGPT), which can only be queried via the aplication programming interface (API) with no access to the parameters. Specifically, for a black-box LLM $f(\cdot)$ that maps any input ${\bf x}_t$ along with a {\it prompt} $\bbtheta$ to a distribution over the language space, a score function $s(\cdot, \cdot)$ will be adopted to evaluate the agreement between $f(\bbtheta, {\bf x}_t)$ and the ground truth $y_t$ in the validation set ${\cal D}_{X,Y}^{ V}$. Then,  the prompt optimization objective is $r(\bbtheta):= \mathbb{E}_{({\bf x}_t,y_t)\sim {\cal D}_{X,Y}^{ V}} s (f(\bbtheta, {\bf x}_t), y_t)$, where $\mathbb{E}$ denotes expectation. This is a black-box optimization task with high evaluation cost, rendering it proper for BO. 

However, directly optimizing over $\bbtheta$ is formidably challenging because $\bbtheta$ here is high-dimensional, and the task is combinatorially complex with complicated structural constraints: in order to be taken by black-box LLMs, the instruction is a combination of discrete tokens that have to comprise human-readable and task-relevant sentence(s). To cope with these challenges, the key idea is to optimize a soft prompt $\boldsymbol{\xi}$ instead, which is fed to a pre-trained white-box LLM $g(\cdot)$ to yield human-readable and task-relevant instruction via in-context learning with $S$ exemplars collected in $\mathcal{E}:=\{({\bf x}_s , y_s )\}_{s=1}^S$ drawn from the target task~\ccite{chen2023instructzero,linuse}. With $\bbtheta := g(\boldsymbol{\xi}, \mathcal{E})$, this combinatorial optimization task is converted to a continuous one. Nevertheless, the resulting problem is still high-dimensional (e.g., the dimension of $\boldsymbol{\xi}$ is thousands for Vicuna) which is challenging to solve. To address this, the solution is to optimize a lower-dimensional vector $\bbphi \in \mathbb{R}^{d'}$ with $d' \ll d$, which will be mapped to a $d$-dimensional space using a random projection matrix $\mathbf{R}\in \mathbb{R}^{d\times d'} $ as $\boldsymbol{\xi}:=\mathbf{R}\bbphi$. Here, each entry in $\mathbf{R}$ is sampled from a normal or uniform distribution. As argued in~\cite{wang2016bayesian}, this random projection is distance-preserving in the sense that BO in the original space and dimension-reduced space are consistent. Further, the powerful in-context learning capability of the white-box LLM can generate expressive and task-relevant instructions given the low-dimensional soft prompt together with the exemplars $\mathcal{E}$. Hence, the prompt optimization task is recast to a low-dimensional continuous one as: $\bbphi^*= \arg\max_{\bbphi \in \mathbb{R}^{d'}}\ r(\bbphi)$, where $r(\bbphi):= \mathbb{E}_{({\bf x}_t,y_t)\sim {\cal D}_{X,Y}^{ V}} s (f(g({\bf R}\bbphi, {\cal }), {\bf x}_t), y_t)$; see also Fig.~\ref{fig:BO}(b) for each evaluation of the soft prompt $\bbphi$. 

To solve this black-box prompt optimization problem in a query-efficient manner, a GP-based BO approach, termed ``InstructZero", was developed in~\cite{chen2023instructzero}. The key idea is to design an interpretable kernel function that can capture the correlation of two output scores through the similarity of the continuous variables in the low-dimensional space. To proceed, the pairwise correlation of two output scores is calculated as ${\bf K}_{i,j} = \mathbb{E}_{{\bf x}_t\sim {\cal D}_{X}^{ V}}\ {\rm sim}(f(\bbtheta_i,{\bf x}_t), f(\bbtheta_j,{\bf x}_t))$, where ${\rm sim}(\cdot, \cdot)$ refers to the similarity between the zero-shot predictions on the target task, e.g., exact match, F1, or BLEU score. Let $l: \mathbb{R}^d \times \mathbb{R}^d\rightarrow \mathbb{R}$ denote the commonly used GP kernel function $l(\bbphi,\bbphi)$ (e.g., Matern or squared exponential) over the low-dimensional continuous soft prompts. Given $n$ evaluated pairs ${\cal D}_n:=\{(\bbphi_{n'}, r_{n'})\}_{n'=1}^n$, the generalized Nystrom extension is leveraged to obtain the kernel function $\kappa(\bbphi,\bbphi'):=\mathbf{l}_n^\top (\bbphi) \mathbf{L}_n^{-1}\mathbf{K}_n \mathbf{l}_n (\bbphi')$~\cite{chen2023instructzero}. This kernel preserves the instruction similarity in the soft prompt space: evaluating $\kappa(\cdot,\cdot)$ on the soft prompts $[\bbphi_1,\ldots, \bbphi_n]$ yields the kernel matrix ${\bf K}$. For a new soft prompt $\bbphi$, this instruction-coupled kernel facilitates smooth extrapolation. Thus, by combining the kernels of the two spaces the proposed kernel aligns BO in the latent space soft prompts with the instruction optimization in the combinatorial and structured space.

Rather than using GPs to model the mapping from $\bbphi$ to $r$, one can alternatively adopt a NN as the surrogate model given its expressiveness in modeling non-stationary functions~\ccite{linuse}. As with InstructZero, a pretrained LLM $g(\cdot)$ is adopted to transform the low-dimensional soft prompt $\bbphi$ to the instruction $\bbtheta$. Then, a multi-layer perceptron is employed to model the mapping from $\bbtheta$ to $r$. The key in this NN mapping is to obtain the uncertainty value $\sigma_n (\bbtheta)$ ($\bbtheta:=g({\bf R}\bbphi, {\cal E})$) from the NN training. The UCB-based AF is leveraged to select the next $\bbphi_{n+1}$ that promotes the {\it exploration-exploitation} trade-off.

\subsection{Challenges and emerging topics for designing AI-based SP education}
While the BO framework bears great potential of automating the design of AI tools for SP education, several emerging topics remain to be addressed before adapting it to different scenarios.  

\noindent 
$\bullet$ \emph{Multi-objective (MO) generalization with adaptivity and robustness}. 
The design of AI-based SP education is guided by multiple performance metrics (e.g., efficacy, cost, or fairness). The goal here is to generalize the aforementioned single-objective BO to coordinate such multiple and possibly conflicting objectives collected in the $B\times 1$ vector ${\bf r}(\bbtheta):=[r_1(\bbtheta), \ldots, r_B(\bbtheta)]^\top$.  Instead of seeking a unique global solution, the goal of such MOBO is to acquire a set of {\it Pareto-optimal} solutions (Pareto frontier) in as small number of evaluations as possible. A point is called Pareto optimal if it cannot be improved in any of the objectives without compromising some other objective.  The first attempt towards finding the Pareto frontier will be to convert these multiple objectives into a single one using judiciously chosen weights $\{w_{n,b}\}_b$ as $\bar{z}_n = \sum_{b=1}^B w_{n,b} z_{n,b}$. For the sake of {\it adaptivity and robustness}, one can model the map from ${\bf x}$ to $\bar{y}_n$ using the recent development of ensemble surrogate models~\cite{lu2023surrogate}~\ccite{polyzos2023bayesian}, and accordingly design the AF based on existing rules~\cite{garnett2023bayesian}. To further incorporate user-specific preference constraints over candidate objectives, novel acquisition rules will be developed building on existing constrained BO setups~\cite{garnett2023bayesian}.

\noindent 
$\bullet$ \emph{Human-machine collaborative design}. While BO-based approaches can automate the design of AI methods without humans in the loop, there might be scenarios that human feedback or designs are available. How to judiciously incorporate such information to accelerate this design process entails to be investigated. In the scenario where the human preference feedback from a pair of designs is available, one can formulate a preferential BO problem, where, instead of one evaluation point in the plain BO, a pair of query points $\{\bbtheta_n, \bbtheta_n^{'}\}$ will be selected by modeling a latent preference function, given by $h(\bbtheta_n,\bbtheta_n^{'}) := r(\bbtheta_n)-r(\bbtheta_n^{'})$. The problem of prompt optimization for black-box LLMs with human preference feedback has been investigated in~\cite{lin2024prompt}, where $r(\bbtheta)$ is modeled by a NN. In addition to the preferential BO framework, we can incorporate prior belief or domain expertise about the property of the objective function at hand, including the prior distribution of the function optimum or optimizer. This belief can be regarded as a form of likelihood which, together with observations, allows one to obtain these samples from the posterior, based on which one can readily form AFs using different rules to acquire the next point. Further, human feedback/interaction can be inserted during the BO learning process. Here, the human feedback can be transformed into constraint sets, which can be further leveraged to guide the hyperparameter selection of the GP model or acquisition design.

\noindent
$\bullet$ \emph{Distributed AI design for SP education}. 
In many cases, multiple universities or organizations want to collaboratively design AI methods for SP education without sharing potentially sensitive information. For example, University of Minnesota, University of California and University of Georgia may want to jointly design LLM-based intelligent SP textbooks. In this context, there are $Q$ local agents, each of which actively collects ${\cal D}_n^q$ so as to optimize a global objective function $r({\bbphi})$. Toward this goal, all local data must be sent to a central server to learn the {\it nonparametric} GP model, which unfortunately incurs severe privacy concerns and high communication overhead. To address this issue, each agent can advocate a parametric estimate of the GP model inspired by the random feature approximation, based on which the local dataset ${\cal D}_n^q$ will be acquired~\ccite{lu2023surrogate}. Instead of exchanging raw data, local agents send statistics of the parametric model to the server for model aggregation, thus preserving raw data privacy. Relying on different data generation assumptions across agents would provide alternative means of aggregating the sought statistics. Besides optimizing the aggregate model, communication-efficient exchanges in the agent-server upload and download operation could be explored. Rather than exchanging posteriors per round, for instance, designing a mechanism at local agents to decide whether information exchange should be initiated, may be a viable strategy. Notably, each agent could use the information reduction of its parameter vector since its last round as a metric. If the measure of information reduction exceeds a threshold, communication would be initiated; otherwise, local agents would continue data collection until the metric exceeds a predetermined threshold.

\section{An example ``Intelligent'' SP textbook} \label{sec:5}

As students increasingly demand more interactive, immersive, and self-paced learning environments, and as the capabilities of the underlying AI tools continue to improve, it is natural to anticipate a gradual shift toward advanced AI-enabled adaptive learning systems. In this section, we discuss a proof-of-concept \emph{intelligent textbook} for SP created with the help of Generative AI using currently-available resources.  The narrative here provides a more broad, high-level overview of the creation process, and we refer the reader to the associated website (\cite{website_sp}, \url{dmccreary.github.io/signal-processing}) for detailed explanations of the individual content creation steps.

\subsection{Intelligent Textbook Features}

\begin{wrapfigure}{r}{0.7\textwidth}
    \centering
    \vspace{-2em}
    \includegraphics[width=0.95\linewidth]{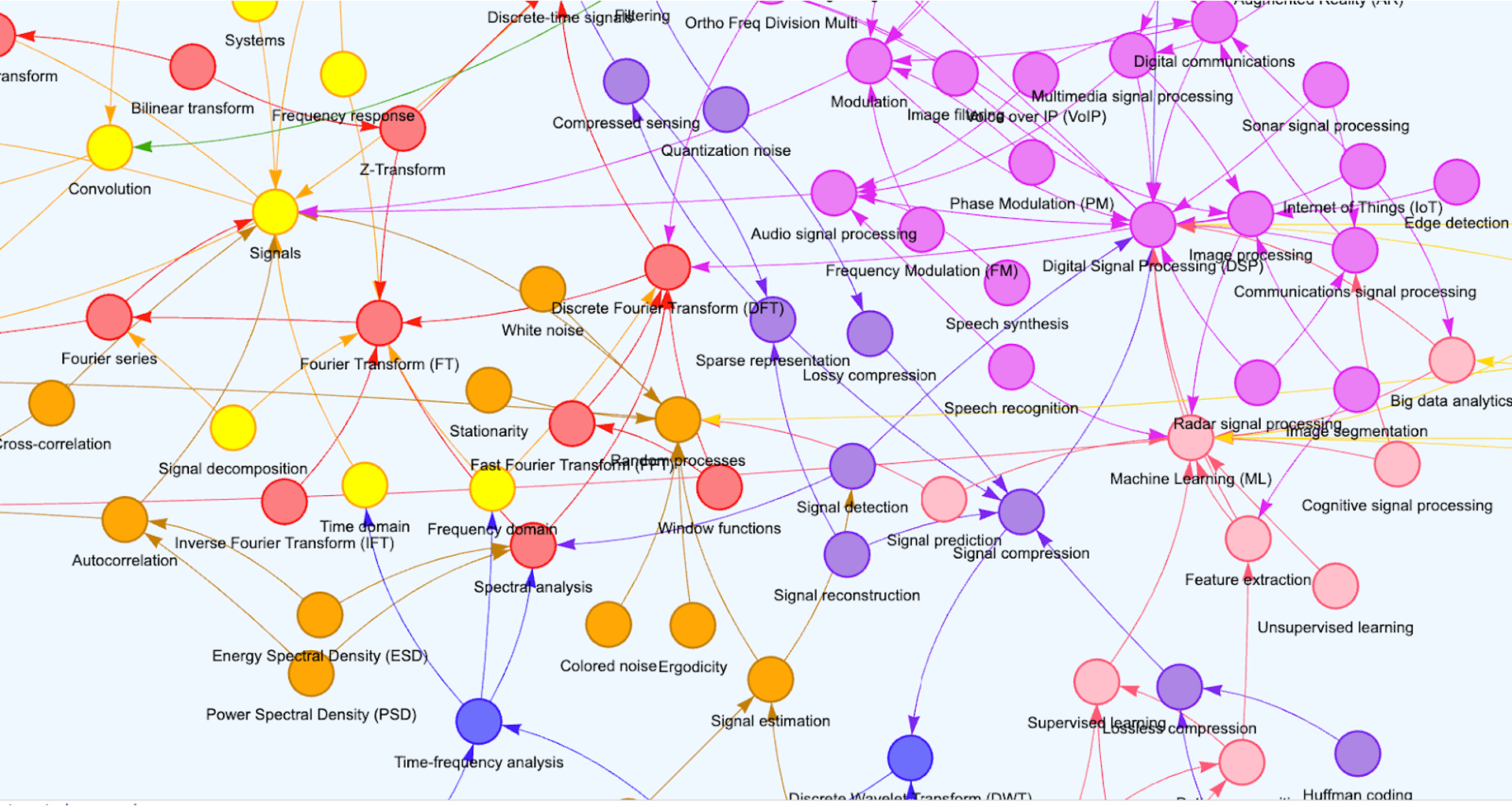}
    \caption{\small A Directed Learning Graph for SP Concepts.  Different colors categorize different concepts; for example, yellow circles represent general concepts and light purple circles indicate more advanced or specialized concepts. }
    \vspace{-2em}
    \label{fig:learninggraph}
\end{wrapfigure}
Similar to the intelligent vehicle classification system that quantifies a range from basic functionality to full autonomy, we envision a framework for intelligent textbooks~\cite{website_level} with varying levels of sophistication.  The simplest incarnations, for instance, would be capable of offering basic features such as keyword search, while more advanced versions would comprise adaptive learning pathways, interactive lessons, and even real-time student guidance. A general-purpose intelligent on-line textbook that can readily adapt to these varying levels could generally incorporate a number of intriguing structural features including, for instance,  a peer-reviewed list of the core SP concepts with tools for customization, a structured \emph{learning graph} illustrating concept dependencies with connections between learning concepts and relevant content (that is, \emph{learning paths}), a search function for efficient content access, and even interactive simulations (\emph{MicroSims}) that can be customized using generative AI.  Stylistically, these resources should nominally also include engaging content that may include, for example, historical timelines and storytelling elements, and could even include up-to-date insights into industry skills required for (SP) professionals. In what follows we describe several of these more innovative AI-based components in more detail.

\subsubsection{Learning Graphs}

Fig.~\ref{fig:learninggraph} depicts an excerpt from an example learning graph designed for the SP course described in \cite{website_sp}. Here, each node represents a concept, while the edges depict dependencies among the concepts. In such representations generally, foundational concepts are positioned on the left, with final course objectives and projects on the right. Paths from concepts to outcomes can thus encode focused learning pathways. Just as learning objectives can vary, so can entire courses -- some may focus on theoretical foundations, for example, while others prioritize hands-on experience with equipment such as oscilloscopes and signal generators. Customized sets of concepts and learning objectives can collectively describe a curated learning experience objective. Next we comment on how these personalized learning objectives are populated with (credible and trusted) content.

\subsubsection{Learning Paths}

Generating high-quality, consistent educational content requires providing AI models with appropriate context. OpenAI and Anthropic have introduced ``Project'' features that allow users to provide a detailed personalized project graph that can be used by the generative AI tools. In other words, once the key concepts in a SP course have been defined, the educator can upload these concepts into the project area, and they will be used as ground truth to anchor the LLM responses, reducing the need for extensive prompt engineering and improving text quality. Further, BO-based prompt optimization approaches, outlined in Sec.~\ref{sec:4} can be leveraged to generate desired course materials in a query-efficient fashion.

\begin{wrapfigure}{r}{0.4\textwidth}
    \centering
    \begin{subfigure}{0.4\textwidth} 
        \centering
        \includegraphics[width=\linewidth]{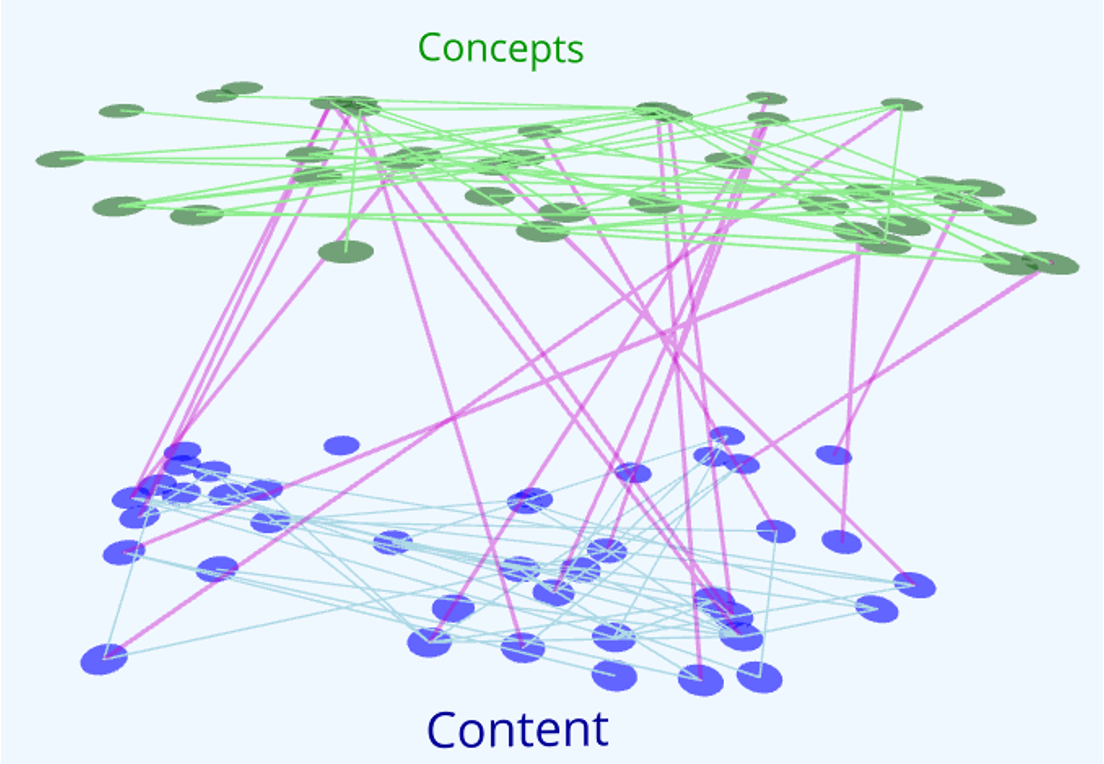}  
        \caption{}
        \label{fig:concept-content-graph}
    \end{subfigure}
    \\
    \begin{subfigure}{0.4\textwidth}
        \centering
        \includegraphics[width=\linewidth]{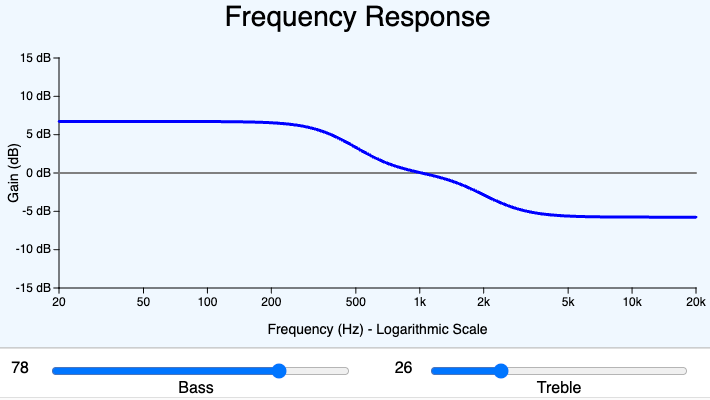}
        \caption{}
        \vspace{-0.5em}
         \label{fig:microsim}
    \end{subfigure}
    \caption{\small Panel (a): Visualization of mapping concepts to content. Panel (b): Example JavaScript Frequency Response MicroSim Widget.}
    \vspace{-2em}
    \label{fig:main}
\end{wrapfigure}

A {\it learning path} is an ordered list of concepts required to achieve a learning objective. Educators often do not expose the underlying concept layer to the students, and students typically engage with content rather than the underlying concepts. But, by separating ``concepts'' from ``content,'' concept lists can be generated using concepts and their dependencies. Figure \ref{fig:concept-content-graph} outlines this idea, where concepts are provided as the abstract/hidden (top, green) layer of vertices that depend on other concepts for their learning order.  The lower level depicts the content related to these concepts. By mapping concept dependencies, AI can 
dynamically generate personalized and coherent content recommendations. AI and embedding techniques can readily automate the connections between concepts and content, creating an adaptive learning experience.

Finally, intelligent textbooks have the ability to consistently monitor student progress by logging their activities with consistent standardized application programming interfaces such as the experience API \emph{xAPI} (\url{xAPI.com}).  Learning events can also be stored in standardized log files such as the Learning Data Store (LDS) standard.  Instructors can then use data science tools to analyze these logs to create better recommendations. Similar to e-commerce recommendation systems, AI can suggest content based on student interactions, e.g. if students rate content or an activity highly, this can be used to influence the content recommendations for next steps in learning. The system dynamically updates recommendations, ensuring students receive content suited to their learning progress, avoiding recommendation of content or concepts that a student has already mastered.

\subsubsection{Micro-Simulations (MicroSims)}

Generative AI facilitates the rapid creation of interactive SP simulations. Today, with simple text-based prompts,  functional JavaScript programs with hundreds of lines of code can be generated directly, often with minimal refinements.  With a library of reference templates and careful prompt engineering, even complex simulations can generated. Fig.~\ref{fig:microsim} depicts an example of a ``widget'' that allows the student to view the changes to a frequency response diagram as (reconfigurable) base and treble controls are modified. As alluded above, one of the striking aspects of these kinds of interactive examples is that they can be created by users who have little to no coding experience, and who may not yet even be fully understand the underlying physical principles being demonstrated.  In this way, such resources truly do enable fully new educational pathways.

\subsection{Prompt Engineering for SP Content Generation}

\begin{wrapfigure}{r}{0.5\textwidth}
    \centering
    \vspace{-1em}
    \includegraphics[width=0.9\linewidth]{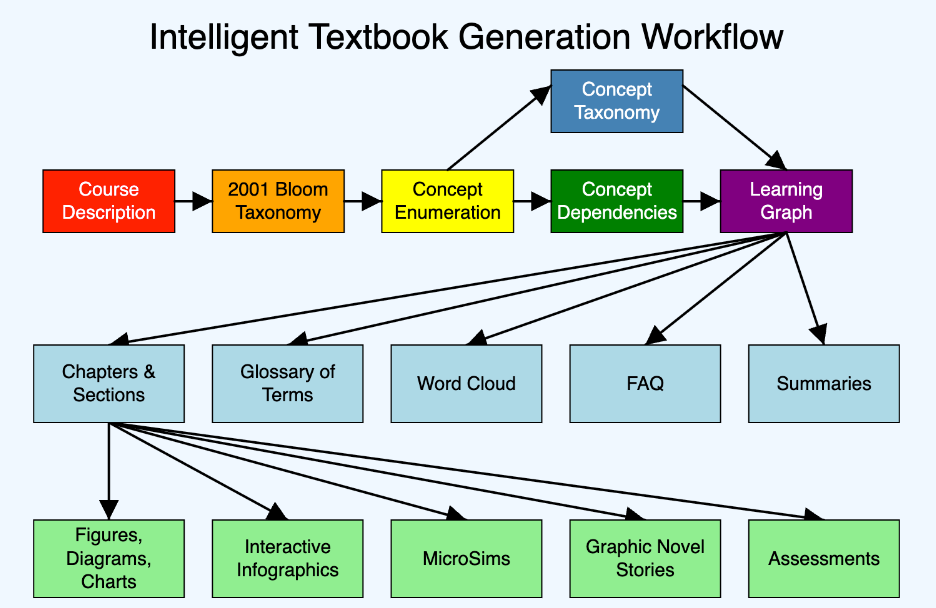}
    \caption{\small Intelligent Textbook Generating Workflow.}
    \vspace{-1em}
    \label{fig:workflow}
\end{wrapfigure}

Much of our work in generating the SP proof of concept is creating a workflow of tasks that move from a high-level course description to generating detailed content, such as the aforementioned MicroSims.  We use a workflow similar to that in Figure~\ref{fig:workflow}. It is critical to understand that LLMs are only a model of language, not an easy-to-query knowledge graph of your course. Intelligent textbook authors create precise, detailed, high-quality, curated, and peer-reviewed learning graphs that can be used to help students efficiently achieve their SP learning objectives. Below we summarize prompt engineering techniques we have found helpful.

\subsubsection{Decomposition}

The primary skill needed to write effective prompts for lengthy textbooks is learning how to write problem decomposition prompts. For example, LLMs cannot generate an entire chapter and all the diagrams in one pass. However, an LLM can create the outline of a chapter and suggest where useful figures and charts can be placed in a chapter. When generating MicroSims, asking an LLM to decompose a MicroSim into a dozen small JavaScript functions is often helpful. For example, you might ask it to develop a JavaScript function to display a waveform's frequency distribution in a chart and then place that chart in a simulation that executes an FFT in real time.

\subsubsection{Providing The Appropriate Context}

SP encompasses a vast range of mathematical concepts, from basic time-domain analysis to advanced machine learning techniques. When crafting prompts for SP applications, providing sufficient context about the specific subdomain in which one is working is essential. For example, one would distinguish between digital SP, analog SP, biomedical SP, or communications SP, as each has unique terminology, assumptions, and methodologies. Similarly, one needs to start their prompts by clearly defining the signal type that is being used, e.g., continuous-time or discrete-time signals, deterministic or stochastic, etc. These fundamental characteristics dramatically influence the appropriate processing techniques and mathematical frameworks.

\subsubsection{Mathematical Precision and Notation}

SP is inherently mathematical, so the prompts should be precise about the mathematical notation and conventions that needs to be displayed in the text. In our case study, we use the LaTeX standards to represent equations, but we find many variations even within the LaTeX standards. So, our prompts also include rules for generating LaTeX and samples of other equations we prefer. For example, after each equation, we use a ``where'' block of text describing each variable in the equation and its definition consistent with the definitions in our glossary of terms.

\subsubsection{Contextualizing Application Domains}

Different application domains in SP have evolved unique vocabularies and standard practices. Audio SP emphasizes perceptual quality metrics, while biomedical SP focuses on physiological relevance and clinical validation. Communications SP prioritizes channel modeling and error rates, while image processing emphasizes spatial relationships and visual quality. Thus, it is important to frame prompts within the appropriate application context, instead of asking generically about ``filtering techniques,'' specify ``low-latency audio filtering for real-time guitar effects processing'' or ``adaptive filtering for ECG artifact removal in noisy clinical environments.'' This contextual framing helps the LLM provide more relevant and practical responses.

\subsubsection{Prompts for Generating Sample Code}

Python has become the \emph{de facto} language of data science, deep learning, and many SP applications. The vast number of open-source Python libraries allows generative AI to quickly create high-quality code that solves complex problems in just a few lines. We have found Python programs to be of higher quality than proprietary languages because LLMs have more sample programs available on the public web with which to be trained. Python also has the added advantage that many LLM tools now include the ability to not only generate Python code, but these tools also execute the code in a virtual machine and return the results of data analysis in graphic images directly in the LLM tool.

\subsubsection{Generating JavaScript MicroSims}

Our SP proof-of-concept textbook uses JavaScript and p5.js (\url{p5js.org}) to create complex interactive charts and execute complex simulations directly within the browser. Using JavaScript makes it easy for students with a web browser on their computer or cell phone to execute our examples. The p5.js library is ideal for generating interactive SP simulations because of its vast code base and consistent support by a wide community of developers.

As of June 2025, Anthropic's Claude web application now executes p5.js examples directly in the development tool that runs in a browser. This direct execution allows one to enter prompts on the left side of the browser and quickly view layout changes without copying and pasting the generated code into the pj.js editing tool. This small change makes it even easier for non-technical staff to generate MicroSims and add interactivity to static diagrams.

\subsubsection{Using Educational Theory in Prompts}

Educational theory provides detailed knowledge about techniques for slowly building a progressive understanding of complex concepts. Great textbooks start with fundamental concepts before moving to advanced topics. When generating lesson plans or requesting code examples or simulations, it is important to specify the intended audience level and what concepts in the learning graph they have recently mastered. Providing this context to an LLM is one of the best ways to create effective learning. It is also helpful to request multiple explanation approaches for complex concepts. Many students learn best when they are first given a metaphor or simulation that they have a concrete understanding of. Prompts that ask for a mix of metaphors, stories, mathematical derivations, and intuitive explanations are helpful. Visual analogies can be particularly powerful for SP concepts—the relationship between time and frequency domains, the concept of convolution, or the meaning of phase relationships.

\subsection{Verification and Validation Strategies}

\begin{wrapfigure}{r}{0.4\textwidth}
    \centering
    \vspace{-1em}
    \includegraphics[width=0.9\linewidth]{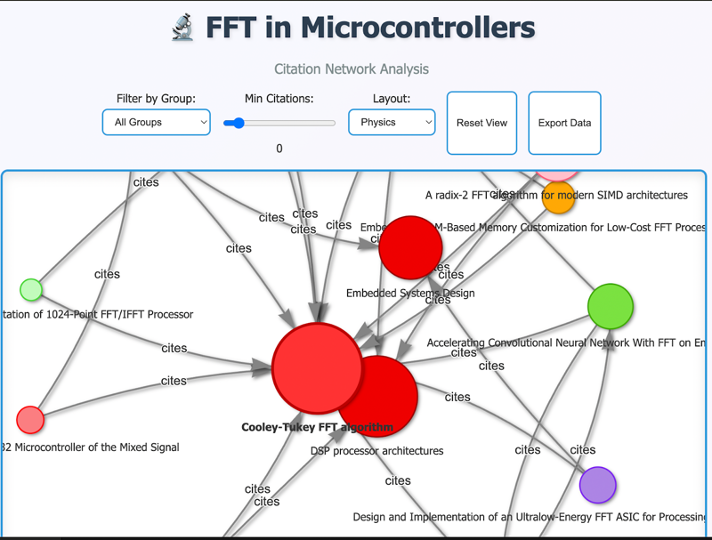}
    \caption{\small Visualization: Reference Veracity of AI Sources.}
    \vspace{-2em}
    \label{fig:pagerank}
\end{wrapfigure}

Although generative AI's task length capabilities double approximately every seven months, there are still no guarantees that generated educational content will be more effective than a well-written static paper textbook. There are techniques one can use various LLMs to cross-check the veracity of your generated content with various LLMs. Cross-checking can be as simple as having one LLM cross-check the content generated by another.  A judicious approach might be to have a first draft of a chapter generated by a smaller LLM such as DeepSeek R1 running, e.g., on a local consumer-grade GPU. One can then upload this content to a larger language model for verification. These larger models have a higher cost per token analyzed. Using established software packages as reference implementations (or to implement intermediate computations) can also help improve the viability of the content produced. MATLAB's SP Toolbox, Python's SciPy, and GNU Radio provide well-tested implementations of standard algorithms. 

\subsection{Ways to Increase the Quality of References and Citations}

One additional challenge that LLMs have had in the past is that they ``fabricated'' plausible citations that do not actually exist. In our proof-of-concept site, we provide examples of generative AI prompts that find the most relevant papers to a specific SP topic and then generate a citation graph for these papers to find the most influential papers. We also demonstrate a MicroSim that allows the use of graph algorithms, such as PageRank~\ccite{page1999pagerank}, to find the most credible papers for a given topic; as in Fig.~\ref{fig:pagerank}. 

\subsection{(Near-)Future Extensions}

While MicroSims are one currently-realizable learning object type that can be dynamically ordered in a knowledge graph for manipulation and updating to enhance learning and teaching, other generative media types could also be used in learning and teaching SP. Interactive \emph{multi-modal} learning objects could include different types of content that may be similarly auto-generated, including:
\begin{enumerate}
    \item {\bf Projections and Prototypes:} 2.5 and 3D models using JavaScript libraries like \emph{three.js} (\url{threejs.org}).
 \item {\bf Sound Simulations:} AI generated sounds to illustrate audio processing and filtering concepts.
 \item {\bf Visual Storytelling:} AI-generated visualizations and interactive plots enhance content engagement, and in the future, will likely even automatically produce curated educational videos.  Our proof of concept site demonstrates the use of Generative AI to create entertaining graphic novel-types stories from the history of SP.
 \item {\bf Hands On Projects:}  We have also extended the concept of intelligent textbooks to include sample hands on projects such as building a spectrum analyzer with a display using a Raspberry Pi Pico 2.
\end{enumerate}
We (again) encourage readers to visit the companion proof-of-concept website \cite{website_sp} for additional details.

\section{Summary} \label{sec:6}

In this paper, we considered the use of generative AI tools in designing and augmenting signal processing education, examining several selected technical challenges and an example practical use case. Perhaps the essential take-away message here should be that while generative AI tools are poised to have a significant impact on signal processing education via various avenues, ongoing foundational work and prudence in practical usages are necessary to address AI's various shortcomings, and in order to ensure its fair, responsible, and trustworthy utilization.

\section*{Acknowledgments}

The authors graciously acknowledge Sharat Batra, Valerie Lockhart, and Troy Peterson for a number of enlightening conversations regarding their current efforts using AI as an educational resource.

\bibliographystyle{IEEEtran}
\bibliography{refs-jdh,ref_Qin,ref-yanning,refs-jia-yue}

\begin{thebibliography}{10}
\providecommand{\url}[1]{#1}
\csname url@samestyle\endcsname
\providecommand{\newblock}{\relax}
\providecommand{\bibinfo}[2]{#2}
\providecommand{\BIBentrySTDinterwordspacing}{\spaceskip=0pt\relax}
\providecommand{\BIBentryALTinterwordstretchfactor}{4}
\providecommand{\BIBentryALTinterwordspacing}{\spaceskip=\fontdimen2\font plus
\BIBentryALTinterwordstretchfactor\fontdimen3\font minus
  \fontdimen4\font\relax}
\providecommand{\BIBforeignlanguage}[2]{{%
\expandafter\ifx\csname l@#1\endcsname\relax
\typeout{** WARNING: IEEEtran.bst: No hyphenation pattern has been}%
\typeout{** loaded for the language `#1'. Using the pattern for}%
\typeout{** the default language instead.}%
\else
\language=\csname l@#1\endcsname
\fi
#2}}
\providecommand{\BIBdecl}{\relax}
\BIBdecl

\bibitem{novik2022role}
V.~Novik, ``The role of learning in returns to college major: evidence from 2.8
  million reviews of 150,000 professors,'' \emph{Available at SSRN 4275668},
  2022.

\bibitem{plass2010cognitive}
J.~L. Plass, R.~Moreno, and R.~Br{\"u}nken, \emph{Cognitive load theory}.\hskip
  1em plus 0.5em minus 0.4em\relax Cambridge University Press, 2010.

\bibitem{de2010cognitive}
T.~D. Jong, ``Cognitive load theory, educational research, and instructional
  design: Some food for thought,'' \emph{Instructional Science}, vol.~38,
  no.~2, pp. 105--134, 2010.

\bibitem{akccayir2018flipped}
G.~Ak{\c{c}}ay{\i}r and M.~Ak{\c{c}}ay{\i}r, ``The flipped classroom: A review
  of its advantages and challenges,'' \emph{Computers \& Education}, vol. 126,
  pp. 334--345, 2018.

\bibitem{bishop2013flipped}
J.~Bishop and M.~A. Verleger, ``The flipped classroom: A survey of the
  research,'' in \emph{2013 ASEE Annual Conference \& Exposition}, 2013, pp.
  23--1200.

\bibitem{kerr2015flipped}
B.~Kerr, ``The flipped classroom in engineering education: A survey of the
  research,'' in \emph{2015 International Conference on Interactive
  Collaborative Learning (ICL)}.\hskip 1em plus 0.5em minus 0.4em\relax IEEE,
  2015, pp. 815--818.

\bibitem{van2013flipping}
B.~V. Veen, ``Flipping signal-processing instruction [sp education],''
  \emph{IEEE Signal Processing Magazine}, vol.~30, no.~6, pp. 145--150, 2013.

\bibitem{wiki:GPT-4}
Wikipedia, ``{GPT-4} --- {W}ikipedia{,} the free encyclopedia,''
  \url{http://en.wikipedia.org/w/index.php?title=GPT-4}, 2025, [Online;
  accessed 28-January-2025].

\bibitem{AIFilms}
\BIBentryALTinterwordspacing
A.~Wanjala. (2024) {AI} tools are making films: {H}ere are 5 sites where you
  can watch. [Online]. Available:
  \url{https://www.makeuseof.com/sites-where-you-can-watch-ai-films/}
\BIBentrySTDinterwordspacing

\bibitem{wang2023scientific}
H.~Wang \emph{et~al.}, ``Scientific discovery in the age of artificial
  intelligence,'' \emph{Nature}, vol. 620, no. 7972, pp. 47--60, 2023.

\bibitem{yang2024leandojo}
K.~Yang \emph{et~al.}, ``Leandojo: {T}heorem proving with retrieval-augmented
  language models,'' \emph{Advances in Neural Information Processing Systems},
  vol.~36, 2024.

\bibitem{jumper2021highly}
J.~Jumper \emph{et~al.}, ``Highly accurate protein structure prediction with
  {A}lpha{F}old,'' \emph{nature}, vol. 596, no. 7873, pp. 583--589, 2021.

\bibitem{mak2024artificial}
K.-K. Mak, Y.-H. Wong, and M.~R. Pichika, ``Artificial intelligence in drug
  discovery and development,'' \emph{Drug discovery and evaluation: {S}afety
  and pharmacokinetic assays}, pp. 1461--1498, 2024.

\bibitem{chowdhury20206g}
M.~Z. Chowdhury, M.~Shahjalal, S.~Ahmed, and Y.~M. Jang, ``6g wireless
  communication systems: {A}pplications, requirements, technologies,
  challenges, and research directions,'' \emph{IEEE Open Journal of the
  Communications Society}, vol.~1, pp. 957--975, 2020.

\bibitem{gillespie2023trust}
N.~Gillespie, S.~Lockey, C.~Curtis, J.~Pool, and A.~Akbari, ``Trust in
  artificial intelligence: A global study,'' \emph{The University of Queensland
  and KPMG Australia}, vol.~10, 2023.

\bibitem{zawacki2019}
O.~Zawacki-Richter, V.~I. Marín, M.~Bond, and F.~Gouverneur, ``Systematic
  review of research on artificial intelligence applications in higher
  education,'' \emph{International Journal of Educational Technology in Higher
  Education}, vol.~16, no.~1, p.~39, 2019.

\bibitem{luckin2018}
R.~Luckin, \emph{Machine learning and human intelligence: The future of
  education for the 21st century}.\hskip 1em plus 0.5em minus 0.4em\relax UCL
  Institute of Education Press, 2018.

\bibitem{holmes2020}
W.~Holmes, M.~Bialik, and C.~Fadel, \emph{Artificial intelligence in education:
  Promises and implications for teaching and learning}.\hskip 1em plus 0.5em
  minus 0.4em\relax Center for Curriculum Redesign, 2020.

\bibitem{koedinger2019}
K.~R. Koedinger and A.~T. Corbett, ``The impact of intelligent tutoring systems
  on student learning,'' \emph{Educational Psychologist}, vol.~41, no.~3, pp.
  183--206, 2019.

\bibitem{pane2017}
J.~F. Pane, E.~D. Steiner, M.~D. Baird, and L.~S. Hamilton, ``How effective is
  online learning? evidence from a large-scale study of k-12 students,'' RAND
  Corporation, Tech. Rep., 2017.

\bibitem{baker2014}
R.~S. Baker and P.~S. Inventado, \emph{Educational data mining and learning
  analytics}.\hskip 1em plus 0.5em minus 0.4em\relax Springer, 2014.

\bibitem{west2019}
D.~M. West, M.~Rhoads, and J.~Choi, ``How artificial intelligence is
  transforming the future of education,'' \emph{Brookings Institution}, 2019.

\bibitem{kusner2017counterfactual}
M.~J. Kusner, J.~Loftus, C.~Russell, and R.~Silva, ``Counterfactual fairness,''
  \emph{Advances in neural information processing systems}, vol.~30, 2017.

\bibitem{dwork2012fairness}
C.~Dwork, M.~Hardt, T.~Pitassi, O.~Reingold, and R.~Zemel, ``Fairness through
  awareness,'' in \emph{Proceedings of the 3rd innovations in theoretical
  computer science conference}, 2012, pp. 214--226.

\bibitem{hardt2016equality}
M.~Hardt, E.~Price, and N.~Srebro, ``Equality of opportunity in supervised
  learning,'' \emph{Advances in neural information processing systems},
  vol.~29, 2016.

\bibitem{selwyn2020}
N.~Selwyn, \emph{Education and technology: Key issues and debates}.\hskip 1em
  plus 0.5em minus 0.4em\relax Bloomsbury Publishing, 2020.

\bibitem{fan2024graph}
W.~Fan \emph{et~al.}, ``Graph machine learning in the era of large language
  models ({LLM}s),'' \emph{arXiv preprint arXiv:2404.14928}, 2024.

\bibitem{li2024graph}
X.~Li \emph{et~al.}, ``Graph learning in the era of llms: A survey from the
  perspective of data, models, and tasks,'' \emph{arXiv preprint
  arXiv:2412.12456}, 2024.

\bibitem{Wu2020}
Z.~Wu, S.~Pan, F.~Chen, G.~Long, C.~Zhang, and P.~S. Yu, ``A comprehensive
  survey on graph neural networks,'' \emph{IEEE Transactions on Neural Networks
  and Learning Systems}, vol.~32, no.~1, pp. 4--24, 2020.

\bibitem{Kipf2017}
T.~N. Kipf and M.~Welling, ``Semi-supervised classification with graph
  convolutional networks,'' \emph{International Conference on Learning
  Representations (ICLR)}, 2017.

\bibitem{zhou2020}
J.~Zhou, G.~Cui, Z.~Zhang, C.~Yang, Z.~Liu, L.~Wang, C.~Li, and M.~Sun, ``Graph
  neural networks: A review of methods and applications,'' \emph{AI Open},
  vol.~1, pp. 57--81, 2020.

\bibitem{ying2019gnnexplainer}
Z.~Ying, D.~Bourgeois, J.~You, M.~Zitnik, and J.~Leskovec, ``Gnnexplainer:
  Generating explanations for graph neural networks,'' \emph{Advances in neural
  information processing systems}, vol.~32, 2019.

\bibitem{kose2023demystifying}
O.~D. Kose and Y.~Shen, ``Demystifying and mitigating bias for node
  representation learning,'' \emph{IEEE Transactions on Neural Networks and
  Learning Systems}, 2023.

\bibitem{kose2021fair}
------, ``Fair contrastive learning on graphs,'' \emph{IEEE Transactions on
  Signal and Processing over Networks}, vol.~8, pp. 475--488, 2022.

\bibitem{kose2022fairnorm}
------, ``Fast\&fair: {T}raining acceleration and bias mitigation for {GNN}s,''
  \emph{Transactions on Machine Learning Research}, 2023.

\bibitem{kose2022fast}
------, ``Fast\&fair: Training acceleration and bias mitigation for gnns,''
  \emph{Transactions on Machine Learning Research}, 2022.

\bibitem{kose2023fairgat}
------, ``Fair{GAT}: Fairness-aware graph attention networks,'' \emph{ACM
  Transactions on Knowledge Discovery from Data}, vol.~18, no.~7, pp. 1--20,
  2024.

\bibitem{kose2023fairnessfilt}
O.~D. Kose, Y.~Shen, and G.~Mateos, ``Fairness-aware graph filter design,''
  \emph{arXiv preprint arXiv:2303.11459}, 2023.

\bibitem{kose2024fairwire}
O.~D. Kose and Y.~Shen, ``Fairwire: Fair graph generation,'' \emph{Advances in
  neural information processing systems}, 2024.

\bibitem{zhu2019dp}
H.~Zhu, X.~Zuo, and M.~Xie, ``{DP-FT}: A differential privacy graph generation
  with field theory for social network data release,'' \emph{IEEE access},
  vol.~7, pp. 164\,304--164\,319, 2019.

\bibitem{buehler2023generative}
M.~J. Buehler, ``Generative pretrained autoregressive transformer graph neural
  network applied to the analysis and discovery of novel proteins,''
  \emph{Journal of Applied Physics}, vol. 134, no.~8, 2023.

\bibitem{ying2009graph}
X.~Ying and X.~Wu, ``Graph generation with prescribed feature constraints,'' in
  \emph{Proc. International Conference on Data Mining (SIAM)}, 2009, pp.
  966--977.

\bibitem{zhao2023graphgpt}
Q.~Zhao, W.~Ren, T.~Li, X.~Xu, and H.~Liu, ``Graphgpt: Graph learning with
  generative pre-trained transformers,'' \emph{arXiv preprint
  arXiv:2401.00529}, 2023.

\bibitem{gui2023challenge}
G.~Gui and O.~Toubia, ``The challenge of using llms to simulate human behavior:
  A causal inference perspective,'' \emph{arXiv preprint arXiv:2312.15524},
  2023.

\bibitem{ma2024causal}
J.~Ma, ``Causal inference with large language model: A survey,'' \emph{arXiv
  preprint arXiv:2409.09822}, 2024.

\bibitem{liu2024large}
X.~Liu \emph{et~al.}, ``Large language models and causal inference in
  collaboration: A comprehensive survey,'' \emph{arXiv preprint
  arXiv:2403.09606}, 2024.

\bibitem{pearl2009}
J.~Pearl, \emph{Causality: Models, reasoning, and inference}.\hskip 1em plus
  0.5em minus 0.4em\relax Cambridge University Press, 2009.

\bibitem{imbens2015}
G.~W. Imbens and D.~B. Rubin, \emph{Causal inference in statistics, social, and
  biomedical sciences}.\hskip 1em plus 0.5em minus 0.4em\relax Cambridge
  University Press, 2015.

\bibitem{giannakis2018topology}
G.~B. Giannakis, Y.~Shen, and G.~V. Karanikolas, ``Topology identification and
  learning over graphs: Accounting for nonlinearities and dynamics,''
  \emph{Proceedings of the IEEE}, vol. 106, no.~5, pp. 787--807, 2018.

\bibitem{shen2017kernel}
Y.~Shen, B.~Baingana, and G.~B. Giannakis, ``Kernel-based structural equation
  models for topology identification of directed networks,'' \emph{IEEE
  Transactions on Signal Processing}, vol.~65, no.~10, pp. 2503--2516, 2017.

\bibitem{zhang2024simulating}
Z.~Zhang \emph{et~al.}, ``Simulating classroom education with llm-empowered
  agents,'' \emph{arXiv preprint arXiv:2406.19226}, 2024.

\bibitem{xu2024large}
H.~Xu, W.~Gan, Z.~Qi, J.~Wu, and P.~S. Yu, ``Large language models for
  education: A survey,'' \emph{arXiv preprint arXiv:2405.13001}, 2024.

\bibitem{maynez-etal-2020-faithfulness}
J.~Maynez, S.~Narayan, B.~Bohnet, and R.~McDonald, ``On faithfulness and
  factuality in abstractive summarization,'' in \emph{Proceedings of the 58th
  Annual Meeting of the Association for Computational Linguistics}.\hskip 1em
  plus 0.5em minus 0.4em\relax Online: Association for Computational
  Linguistics, Jul. 2020, pp. 1906--1919.

\bibitem{perkovic2024hallucinations}
G.~Perkovi{\'c}, A.~Drobnjak, and I.~Boti{\v{c}}ki, ``Hallucinations in llms:
  Understanding and addressing challenges,'' in \emph{2024 47th MIPRO ICT and
  Electronics Convention (MIPRO)}.\hskip 1em plus 0.5em minus 0.4em\relax IEEE,
  2024, pp. 2084--2088.

\bibitem{tang2024prioritizing}
X.~Tang \emph{et~al.}, ``Prioritizing safeguarding over autonomy: Risks of llm
  agents for science,'' \emph{arXiv preprint arXiv:2402.04247}, 2024.

\bibitem{rossi2024problems}
L.~Rossi, K.~Harrison, and I.~Shklovski, ``The problems of llm-generated data
  in social science research,'' \emph{Sociologica}, vol.~18, no.~2, pp.
  145--168, 2024.

\bibitem{dong2023future}
\BIBentryALTinterwordspacing
Y.~Dong, ``What future do we want with artificial intelligence?'' 2023,
  accessed: 2025-02-10. [Online]. Available:
  \url{https://www.youtube.com/watch?v=w5fmMcrwk_I}
\BIBentrySTDinterwordspacing

\bibitem{su2023unlocking}
J.~Su and W.~Yang, ``Unlocking the power of chatgpt: A framework for applying
  generative ai in education,'' \emph{ECNU Review of Education}, vol.~6, no.~3,
  pp. 355--366, 2023.

\bibitem{espark_ai_education}
\BIBentryALTinterwordspacing
{eSpark Learning}, ``Ai in education: The problem with hallucinations,'' 2023,
  accessed: 2024-11-06. [Online]. Available:
  \url{https://www.esparklearning.com/blog/ai-in-education-the-problem-with-hallucinations/}
\BIBentrySTDinterwordspacing

\bibitem{filippova-2020-controlled}
K.~Filippova, ``Controlled hallucinations: Learning to generate faithfully from
  noisy data,'' in \emph{Findings of the Association for Computational
  Linguistics: EMNLP 2020}.\hskip 1em plus 0.5em minus 0.4em\relax Online:
  Association for Computational Linguistics, Nov. 2020, pp. 864--870.

\bibitem{cao2021hallucinated}
M.~Cao, Y.~Dong, and J.~C.~K. Cheung, ``Hallucinated but factual! inspecting
  the factuality of hallucinations in abstractive summarization,'' \emph{arXiv
  preprint arXiv:2109.09784}, 2021.

\bibitem{longpre2021entity}
S.~Longpre, K.~Perisetla, A.~Chen, N.~Ramesh, C.~DuBois, and S.~Singh,
  ``Entity-based knowledge conflicts in question answering,'' in
  \emph{Proceedings of the 2021 Conference on Empirical Methods in Natural
  Language Processing}, 2021.

\bibitem{2023halluciation}
Z.~Ji \emph{et~al.}, ``Survey of hallucination in natural language
  generation,'' \emph{ACM Comput. Surv.}, vol.~55, no.~12, Mar. 2023.

\bibitem{10.1145/3703155}
L.~Huang \emph{et~al.}, ``A survey on hallucination in large language models:
  Principles, taxonomy, challenges, and open questions,'' \emph{ACM Trans. Inf.
  Syst.}, vol.~43, no.~2, Jan. 2025.

\bibitem{li2023batgptbidirectionalautoregessivetalker}
\BIBentryALTinterwordspacing
Z.~Li, S.~Zhang, H.~Zhao, Y.~Yang, and D.~Yang, ``Batgpt: A bidirectional
  autoregessive talker from generative pre-trained transformer,'' 2023.
  [Online]. Available: \url{https://arxiv.org/abs/2307.00360}
\BIBentrySTDinterwordspacing

\bibitem{liu-etal-2024-lost}
N.~F. Liu \emph{et~al.}, ``Lost in the middle: How language models use long
  contexts,'' \emph{Transactions of the Association for Computational
  Linguistics}, vol.~12, pp. 157--173, 2024.

\bibitem{shi2024incontext}
\BIBentryALTinterwordspacing
W.~Shi \emph{et~al.}, ``In-context pretraining: Language modeling beyond
  document boundaries,'' in \emph{The Twelfth International Conference on
  Learning Representations}, 2024. [Online]. Available:
  \url{https://openreview.net/forum?id=LXVswInHOo}
\BIBentrySTDinterwordspacing

\bibitem{wei2024simplesyntheticdatareduces}
J.~Wei, D.~Huang, Y.~Lu, D.~Zhou, and Q.~V. Le, ``Simple synthetic data reduces
  sycophancy in large language models,'' 2024.

\bibitem{ram-etal-2023-context}
O.~Ram \emph{et~al.}, ``In-context retrieval-augmented language models,''
  \emph{Transactions of the Association for Computational Linguistics},
  vol.~11, pp. 1316--1331, 2023.

\bibitem{baek-etal-2023-knowledge}
J.~Baek, A.~F. Aji, and A.~Saffari, ``Knowledge-augmented language model
  prompting for zero-shot knowledge graph question answering,'' in
  \emph{Proceedings of the 1st Workshop on Natural Language Reasoning and
  Structured Explanations (NLRSE)}.\hskip 1em plus 0.5em minus 0.4em\relax
  Toronto, Canada: Association for Computational Linguistics, Jun. 2023, pp.
  78--106.

\bibitem{wen-etal-2024-mindmap}
Y.~Wen, Z.~Wang, and J.~Sun, ``{M}ind{M}ap: Knowledge graph prompting sparks
  graph of thoughts in large language models,'' in \emph{Proceedings of the
  62nd Annual Meeting of the Association for Computational Linguistics (Volume
  1: Long Papers)}.\hskip 1em plus 0.5em minus 0.4em\relax Bangkok, Thailand:
  Association for Computational Linguistics, Aug. 2024, pp. 10\,370--10\,388.

\bibitem{he2022rethinkingretrievalfaithfullarge}
\BIBentryALTinterwordspacing
H.~He, H.~Zhang, and D.~Roth, ``Rethinking with retrieval: Faithful large
  language model inference,'' 2022. [Online]. Available:
  \url{https://arxiv.org/abs/2301.00303}
\BIBentrySTDinterwordspacing

\bibitem{trivedi-etal-2023-interleaving}
H.~Trivedi, N.~Balasubramanian, T.~Khot, and A.~Sabharwal, ``Interleaving
  retrieval with chain-of-thought reasoning for knowledge-intensive multi-step
  questions,'' in \emph{Proceedings of the 61st Annual Meeting of the
  Association for Computational Linguistics (Volume 1: Long Papers)}.\hskip 1em
  plus 0.5em minus 0.4em\relax Toronto, Canada: Association for Computational
  Linguistics, Jul. 2023, pp. 10\,014--10\,037.

\bibitem{yao2022react}
S.~Yao \emph{et~al.}, ``React: Synergizing reasoning and acting in language
  models,'' \emph{arXiv preprint arXiv:2210.03629}, 2022.

\bibitem{shao-etal-2023-enhancing}
Z.~Shao \emph{et~al.}, ``Enhancing retrieval-augmented large language models
  with iterative retrieval-generation synergy,'' in \emph{Findings of the
  Association for Computational Linguistics: EMNLP 2023}.\hskip 1em plus 0.5em
  minus 0.4em\relax Singapore: Association for Computational Linguistics, Dec.
  2023, pp. 9248--9274.

\bibitem{jiang-etal-2023-active}
Z.~Jiang \emph{et~al.}, ``Active retrieval augmented generation,'' in
  \emph{Proceedings of the 2023 Conference on Empirical Methods in Natural
  Language Processing}.\hskip 1em plus 0.5em minus 0.4em\relax Singapore:
  Association for Computational Linguistics, Dec. 2023, pp. 7969--7992.

\bibitem{banerjee2024llmshallucinateneedlive}
\BIBentryALTinterwordspacing
S.~Banerjee, A.~Agarwal, and S.~Singla, ``{LLM}s will always hallucinate, and
  we need to live with this,'' 2024. [Online]. Available:
  \url{https://arxiv.org/abs/2409.05746}
\BIBentrySTDinterwordspacing

\bibitem{shi-etal-2024-trusting}
W.~Shi, X.~Han, M.~Lewis, Y.~Tsvetkov, L.~Zettlemoyer, and W.-T. Yih,
  ``Trusting your evidence: Hallucinate less with context-aware decoding,'' in
  \emph{Proceedings of the 2024 Conf. NAACL: Human Language Technologies
  (Volume 2: Short Papers)}.\hskip 1em plus 0.5em minus 0.4em\relax Mexico
  City, Mexico: Association for Computational Linguistics, Jun. 2024, pp.
  783--791.

\bibitem{gao-etal-2023-rarr}
L.~Gao \emph{et~al.}, ``{RARR}: Researching and revising what language models
  say, using language models,'' in \emph{Proceedings of the 61st Annual Meeting
  of the Association for Computational Linguistics (Volume 1: Long
  Papers)}.\hskip 1em plus 0.5em minus 0.4em\relax Toronto, Canada: Association
  for Computational Linguistics, Jul. 2023, pp. 16\,477--16\,508.

\bibitem{lei2023chainnaturallanguageinference}
\BIBentryALTinterwordspacing
D.~Lei \emph{et~al.}, ``Chain of natural language inference for reducing large
  language model ungrounded hallucinations,'' 2023. [Online]. Available:
  \url{https://arxiv.org/abs/2310.03951}
\BIBentrySTDinterwordspacing

\bibitem{10.5555/3454287.3454806}
Z.~Yang, T.~Luong, R.~Salakhutdinov, and Q.~Le, \emph{Mixtape: breaking the
  softmax bottleneck efficiently}.\hskip 1em plus 0.5em minus 0.4em\relax Red
  Hook, NY, USA: Curran Associates Inc., 2019.

\bibitem{branco-etal-2021-shortcutted}
R.~Branco, A.~Branco, J.~Ant{\'o}nio~Rodrigues, and J.~R. Silva, ``Shortcutted
  commonsense: Data spuriousness in deep learning of commonsense reasoning,''
  in \emph{Proceedings of the 2021 Conf. on Empirical Methods in Natural
  Language Processing}.\hskip 1em plus 0.5em minus 0.4em\relax Online and Punta
  Cana, Dominican Republic: Association for Computational Linguistics, Nov.
  2021, pp. 1504--1521.

\bibitem{wang-etal-2023-scott}
P.~Wang, Z.~Wang, Z.~Li, Y.~Gao, B.~Yin, and X.~Ren, ``{SCOTT}: Self-consistent
  chain-of-thought distillation,'' in \emph{Proceedings of the 61st Annual
  Meeting of the Association for Computational Linguistics (Volume 1: Long
  Papers)}.\hskip 1em plus 0.5em minus 0.4em\relax Toronto, Canada: Association
  for Computational Linguistics, Jul. 2023, pp. 5546--5558.

\bibitem{paul-etal-2024-making}
D.~Paul, R.~West, A.~Bosselut, and B.~Faltings, ``Making reasoning matter:
  Measuring and improving faithfulness of chain-of-thought reasoning,'' in
  \emph{Findings of the Association for Computational Linguistics: EMNLP
  2024}.\hskip 1em plus 0.5em minus 0.4em\relax Miami, Florida, USA:
  Association for Computational Linguistics, Nov. 2024, pp. 15\,012--15\,032.

\bibitem{xu-etal-2024-faithful}
J.~Xu \emph{et~al.}, ``Faithful logical reasoning via symbolic
  chain-of-thought,'' in \emph{Proceedings of the 62nd Annual Meeting of the
  Association for Computational Linguistics (Volume 1: Long Papers)}.\hskip 1em
  plus 0.5em minus 0.4em\relax Bangkok, Thailand: Association for Computational
  Linguistics, Aug. 2024, pp. 13\,326--13\,365.

\bibitem{albahli2024efficient}
S.~Albahli, ``Efficient hyperparameter tuning for predicting student
  performance with {B}ayesian optimization,'' \emph{Multimedia tools and
  applications}, vol.~83, no.~17, pp. 52\,711--52\,735, 2024.

\bibitem{liu2023pre}
P.~Liu, W.~Yuan, J.~Fu, Z.~Jiang, H.~Hayashi, and G.~Neubig, ``Pre-train,
  prompt, and predict: {A} systematic survey of prompting methods in natural
  language processing,'' \emph{ACM Computing Surveys}, vol.~55, no.~9, pp.
  1--35, 2023.

\bibitem{garnett2023bayesian}
R.~Garnett, \emph{Bayesian {O}ptimization}.\hskip 1em plus 0.5em minus
  0.4em\relax Cambridge University Press, 2023.

\bibitem{Rasmussen2006gaussian}
C.~E. Rasmussen and C.~K. Williams, \emph{Gaussian processes for machine
  learning}.\hskip 1em plus 0.5em minus 0.4em\relax MIT Press Cambridge, MA,
  2006.

\bibitem{chen2023instructzero}
L.~Chen, J.~Chen, T.~Goldstein, H.~Huang, and T.~Zhou, ``Instructzero:
  {E}fficient instruction optimization for black-box large language models,''
  \emph{arXiv preprint arXiv:2306.03082}, 2023.

\bibitem{linuse}
X.~Lin \emph{et~al.}, ``Use your {INSTINCT}: {INST}ruction optimization for
  {LLM}s us{I}ng {N}eural bandits coupled with {T}ransformers,'' \emph{Intl.
  Conf. on Machine Learn.}, 2024.

\bibitem{wang2016bayesian}
Z.~Wang, F.~Hutter, M.~Zoghi, D.~Matheson, and N.~De~Feitas, ``Bayesian
  optimization in a billion dimensions via random embeddings,'' \emph{Journal
  of Artificial Intelligence Research}, vol.~55, pp. 361--387, 2016.

\bibitem{lu2023surrogate}
Q.~Lu, K.~D. Polyzos, B.~Li, and G.~B. Giannakis, ``Surrogate modeling for
  {B}ayesian optimization beyond a single {G}aussian process,'' \emph{IEEE
  Trans. Pattern Anal. Mach. Intell.}, vol.~45, no.~9, pp. 11\,283--11\,296,
  2023.

\bibitem{polyzos2023bayesian}
K.~D. Polyzos, Q.~Lu, and G.~B. Giannakis, ``Bayesian optimization with
  ensemble learning models and adaptive expected improvement,'' in \emph{IEEE
  Intl. Conf. on Acoustics, Speech and Signal Process.}, 2023, pp. 1--5.

\bibitem{lin2024prompt}
X.~Lin, Z.~Dai, A.~Verma, S.-K. Ng, P.~Jaillet, and B.~K.~H. Low, ``Prompt
  optimization with human feedback,'' \emph{arXiv preprint arXiv:2405.17346},
  2024.

\bibitem{website_sp}
D.~McCreary, ``Signal processing with {AI},''
  \url{dmccreary.github.io/signal-processing/}, 2025, [Online; accessed
  6-June-2025].

\bibitem{website_level}
------, ``Five levels of intelligent textbooks,''
  \url{dmccreary.medium.com/five-levels-of-intelligent-textbooks-b81a4c1525a0},
  2025, [Online; accessed 6-June-2025].

\bibitem{page1999pagerank}
L.~Page, S.~Brin, R.~Motwani, and T.~Winograd, ``The pagerank citation ranking:
  Bringing order to the web.'' Stanford Infolab, Tech. Rep., 1999.

\end{thebibliography}


\end{document}